\documentclass[aos,MSNbibl,citesort,seceqn,dvips]{arximspdf}
\usepackage{dcolumn}
\usepackage{graphicx}

\doi{10.1214/12-AOS978} 
\volume{40}
\issue{2}
\pubyear{2012}
\firstpage{785}
\lastpage{811}

\makeatletter
\newcolumntype{d}[1]{D{.}{.}{#1}}
\newcommand{\ve}[1]{{\bolds{#1}}}
\newtheorem{theorem}{Theorem}
\newproclaim{example}{Example}
\newtheorem{proposition}{Proposition}
\newproclaim{assumption}{Assumption}
\newproclaim{definition}{Definition}
\newproclaim{remarks}{Remarks}

\def\endsupplement{%
  \@ifundefined{@sname}{}{\@ifundefined{supp@label}{}{\hypertarget{\supp@label}{}\label{\supp@label}}\textbf{\@sname: }\\}%
  \@ifundefined{@stitle}{}{\textbf{\@stitle\ }}%
  (%
  \@ifundefined{slink@doi}{}{\slink@doi@fmt}%
  \@ifundefined{@sdatatype}{}{; \@sdatatype}%
  \@ifundefined{@sfilename}{}{\@sfilename}%
  ).
  \@ifundefined{@sdescription}{}{\@sdescription}%
  \par}

\makeatother

\begin{document}
\begin{frontmatter}

\title{Perturbation and scaled Cook's distance}
\runtitle{Perturbation and scaled Cook's distance}

\begin{aug}
\author[A]{\fnms{Hongtu} \snm{Zhu}\corref{}\ead[label=e1]{hzhu@bios.unc.edu}\thanksref{t1}},
\author[A]{\fnms{Joseph G.} \snm{Ibrahim}\ead[label=e2]{ibrahim@bios.unc.edu}\thanksref{t1}}
\and
\author[A]{\fnms{Hyunsoon} \snm{Cho}\ead[label=e3]{hyunsoon.cho@nih.gov}}
\thankstext{t1}{Supported by NSF Grant BCS-08-26844 and NIH
Grants RR025747-01, P01CA142538-01, MH086633, EB005149-01 and  AG033387.}
\runauthor{H. Zhu, J. G. Ibrahim and H. Cho}
\affiliation{University of North Carolina at Chapel Hill}
\address[A]{Department of Biostatistics\\
University of North Carolina at Chapel Hill\\
Chapel Hill, North Carolina 27599-7420\\
USA\\
\printead{e1}\\
\phantom{E-mail: }\printead*{e2}\\
\phantom{E-mail: }\printead*{e3}} 
\end{aug}

\received{\smonth{7} \syear{2011}}
\revised{\smonth{1} \syear{2012}}

%
\begin{abstract}
Cook's distance [\textit{Technometrics} \textbf{19} (1977) 15--18]
is one of the most important diagnostic tools for
detecting influential individual or subsets of observations in linear
regression for cross-sectional data.
However, for many complex data structures (e.g.,
longitudinal data), no rigorous approach has been developed to address
a fundamental issue:
deleting subsets with different numbers of observations introduces
different degrees of perturbation to the current model fitted to the
data, and
the magnitude of Cook's distance is associated with the degree of the
perturbation.
The aim of this paper is to address this issue in general parametric
models with complex data structures. We propose a new quantity for
measuring the degree of the perturbation introduced by deleting a subset.
We use stochastic ordering to quantify the stochastic relationship
between the degree of the perturbation and the magnitude of
Cook's distance. We develop several scaled Cook's distances
to resolve the comparison of Cook's distance for different subset deletions.
Theoretical and numerical examples are examined to highlight the broad
spectrum of applications of these scaled Cook's distances in a formal
influence analysis.
\end{abstract}

%

\end{frontmatter}

\section{Introduction}\label{sec1}

Influence analysis assesses
whether a modification of a statistical analysis, called a perturbation
(see Section~\ref{sec2.2}),
seriously affects specific key inferences, such as parameter estimates.
Such perturbation schemes include the deletion of an individual or a
subset of observations, case weight perturbation and covariate
perturbation, among many others \cite{Cook1977,Cook1986,Zhu-etal2007}.
For example, for linear models, a perturbation measures the effect on
the model of deleting a
subset of the data matrix. In general, perturbation measures do not
depend on the data directly,
but rather on its structure via the model.
If a small perturbation has a small effect on the analysis, our
analysis is relatively stable, while
if a large perturbation has a~small effect on the analysis, we learn
that our analysis is
robust~\cite{Critchley-etal2001,Huber1981}. If a small perturbation
seriously influences key results of the analysis, we want to know the
cause~\cite{Cook1986,Critchley-etal2001}.
For instance,
in influence analysis, a set of observations is flagged as
``influential'' if its removal from the dataset produces a
significant difference in the parameter estimates or, equivalently, a~large value of Cook's distance for the current statistical model~\cite{Cook1977,Beckman1983}.

Since the seminal
work of Cook~\cite{Cook1977} on Cook's distance in linear regression for
cross-sectional data, considerable research has been devoted to
developing Cook's distance for detecting
influential observations (or clusters) in more complex data structures
under various statistical models
\cite{Cook1977,Cook-Weisberg1982,Chatterjee-Hadi1988,Andersen1992,Davison-Tsai1992,Wei1998,Haslett1999,Zhu-etal2001,Fung-etal2002}. For example, for
longitudinal data,
Preisser and Qaqish~\cite{Preisser-Qaqish1996} developed Cook's distance for
generalized estimating equations, while Christensen, Pearson and Johnson~\cite{Christensen-etal92},
Banerjee and Frees~\cite{Banerjee-Frees1997} and Banerjee~\cite{Banerjee1998} considered case
deletion and subject
deletion diagnostics for linear mixed models. Furthermore, in the
presence of missing data,
Zhu et al.~\cite{Zhu-etal2001} developed
deletion diagnostics for a large class of statistical models with
missing data. Cook's distance
has been widely used in statistical practice and can be
calculated in popular statistical software, such as SAS and R.

A major research problem regarding Cook's distance that has been
largely neglected in the
existing literature is the development of Cook's distance for
general statistical models with more complex data structures.
The fundamental issue that arises here is that
the magnitude of Cook's distance is positively associated with the
amount of perturbation
to the current model introduced by deleting a subset of observations.
Specifically, a
large value of Cook's distance can be caused by deleting a subset with
a larger number of observations
and/or other causes such as the presence of influential observations in
the deleted subset.
To delineate the cause of a large Cook's distance for a specific subset,
it is more useful to compute Cook's distance relative to the degree of
the perturbation introduced by
deleting the subset~\cite{Critchley-etal2001,Zhu-etal2007}.


%

The aim of this paper is to address this fundamental issue of Cook's
distance for complex data structures in general parametric
models. The main contributions of this paper are summarized as follows:

(a.1) We propose a quantity to measure the degree of perturbation
introduced by deleting a subset in general parametric models.
This quantity satisfies several attractive properties including
uniqueness, nonnegativity, monotonicity and additivity.

(a.2) We use stochastic ordering to quantify the relationship
between the degree of the perturbation and the magnitude of
Cook's distance. Particularly, in linear regression for
cross-sectional data, we first show the stochastic relationship between
the Cook's distances
for any two subsets with possibly different numbers of observations.

(a.3) We develop several scaled Cook's distances and their first-order
approximations
in order to compare Cook's distance for deleted subsets with different
numbers of
observations.


The rest of the paper is organized as follows.
In Section~\ref{sec2}, we quantify the degree
of the perturbation for set deletion and delineate the stochastic
relationship between
Cook's distance and the degree of perturbation. We develop several
scaled Cook's distances and derive their first-order approximations.
In Section~\ref{sec3}, we analyze simulated data and a real dataset using the
scaled Cook's distances. We
give some final remarks in Section~\ref{sec4}.

\section{Scaled Cook's distance}\label{sec2}

\subsection{Cook's distance}\label{sec2.1}

Consider the probability function of a
random vector $\mathbf{Y}^T=({Y}_1^T, \ldots, {Y}_n^T)$, denoted by
$p(\mathbf{Y}|{\ve\theta})$,
where ${\ve\theta}=(\theta_1, \ldots, \theta_q)^T$ is a $q\times1$
vector in an open subset ${\Theta}$ of $R^q$ and $Y_i=(y_{i, 1},
\ldots, y_{i, m_i})$, in which the dimension of~$Y_i$, denoted by
$m_i$, may vary significantly across all $i$.
Cook's distance and many other deletion diagnostics measure
the distance between the maximum likelihood estimators of $\ve\theta$
with and without $Y_i$~\cite{Cook-Weisberg1982,Cook1977}. A
subscript ``[I]'' denotes the relevant quantity with all observations
in $I$ deleted.
Let $\mathbf{Y}_{[I]}$ be a
subsample of $\mathbf{Y}$, with $\mathbf{Y}_I=
\{Y_{(i, j)}\dvtx  (i, j)\in I\}$ deleted, and $p(\mathbf{Y}_{[I]}|{\ve\theta
})$ be its probability function.
We
define the maximum likelihood estimators of ${\ve\theta}$ for the full
sample $\mathbf{Y}$ and a subsample $\mathbf{Y}_{[I]}$ as
%
\begin{equation}\label{BayesianLOCAL3}
\hat{{\ve\theta}}=\mathop{\operatorname{argmax}}_{\ve\theta} \log p(\mathbf{
Y}|{\ve\theta}) \quad \mbox{and}\quad   \hat{{\ve\theta}}_{[I]}=\mathop{\operatorname{argmax}}_{\ve
\theta} \log
p\bigl(\mathbf{Y}_{[I]}|{\ve\theta}\bigr),
\end{equation}
respectively.
Cook's distance for $I$, denoted by $\operatorname{CD}(I)$, can be defined as
follows:
%
\begin{equation}\label{BayesianLOCAL4}
\operatorname{CD}(I)=\bigl(\hat{{\ve\theta}}_{[I]}-\hat{{\ve\theta}}\bigr)^T
{G}_{n\theta}\bigl(\hat{{\ve\theta}}_{[I]}-\hat{{\ve\theta}}\bigr),
\end{equation}
where ${G}_{n\theta}$ is
chosen to be a positive definite matrix.
The matrix ${G}_{n\theta}$ is not changed or re-estimated when a
subset of the data is deleted.
Throughout the paper, ${G}_{n\theta}$
is set as $ -\partial^2_\theta\log p(\mathbf{Y}|\hat{\ve\theta})$ or its
expectation,
where $\partial_\theta^2$
represents the second-order derivative with respect to $\ve\theta$.
For clustered data, the observations within the same cluster are
correlated. A sensible model $p(\mathbf{Y}|{\ve\theta})$ should explicitly
model the
correlation structure in the clustered data
and thus $ -\partial^2_\theta\log p(\mathbf{Y}|\hat{\ve\theta})$
implicitly incorporates such a correlation structure.

More generally, suppose that one is interested in a subset of $\ve\theta
$ or $q_1$ linearly independent combinations of $\ve\theta$, say $\mathbf{
L}^T\ve\theta$,
where $\mathbf{L}$ is a $q\times q_1$ matrix with rank$(\mathbf{L})=q_1$
\cite{Banerjee-Frees1997,Cook-Weisberg1982}.
The partial influence of the subset $I$ on $\mathbf{L}^T\hat{{\ve\theta
}}$, denoted by $\operatorname{CD}(I|\mathbf{L})$, can be defined as
%
\begin{equation}
\operatorname{CD}(I|\mathbf{L})=\bigl(\hat{{\ve\theta}}_{[I]}-\hat{{\ve\theta}}\bigr)^T
\mathbf{L}\{\mathbf{L}^T{G}_{n\theta}^{-1}\mathbf{L}\}^{-1}\mathbf{L}^T\bigl(\hat{{\ve
\theta}}_{[I]}-\hat{{\ve\theta}}\bigr).
\end{equation}
For notational simplicity, even though we may focus on a subset of $\ve
\theta$, we do not distinguish between $\operatorname{CD}(I|\mathbf{L})$ and
$\operatorname{CD}(I)$ throughout the paper.\vadjust{\goodbreak}

Based on (\ref{BayesianLOCAL4}), we know
that Cook's distance $\operatorname{CD}(I)$ is explicitly determined by three
components, including
the current model fitted to the data, denoted by ${\mathcal M}$, the
dataset $\mathbf{Y}$ and the subset $I$, itself.
Cook's distance is also implicitly determined by the goodness of fit
of ${\mathcal M}$ to $\mathbf{Y}$ for $I$, denoted by $G(I|\mathbf{Y},
{\mathcal M})$, and
the degree of the perturbation to ${\mathcal M}$ introduced by
deleting the subset $I$, denoted by ${\mathcal P}(I|{\mathcal M})$.
Thus, we may
represent $\operatorname{CD}(I)$ as
follows:
%
\begin{equation}\label{size1}
\operatorname{CD}(I)=F_1(I, {\mathcal M}, \mathbf{Y})=F_2({\mathcal
P}(I|{\mathcal M}), G(I| \mathbf{Y}, {\mathcal M})),
\end{equation}
where $F_1(\cdot, \cdot, \cdot)$ and $F_2(\cdot, \cdot)$ represent
nonlinear functions.

We may use the
value of $ \operatorname{CD}(I)$ to assess the influential level of the
subset~$I$.
We may regard a subset $I$ as influential if either the value of
$\operatorname{CD}(I)$ is relatively large, compared with other
Cook's distances, or the magnitude of $\operatorname{CD}(I)$ is greater than
the critical points of the $\chi^2$ distribution~\cite{Cook-Weisberg1982}.
However, for complex data structures, we will show that it is useful to
compare Cook's distance relative to its
associated degree of perturbation.

\subsection{Degree of perturbation}\label{sec2.2}

Consider the subset ${I}$ and the current mod\-el~${\mathcal M}$.
We are interested in developing a measure to quantify
the degree of the perturbation to ${\mathcal M}$ introduced by deleting
the subset $I$, regardless of the observed data $\mathbf{Y}$.
We emphasize here that the degree of perturbation is a~property of
the model, unlike Cook's distance which is also a property of $\mathbf{Y}$.
Abstractly,
${\mathcal P}(I|{\mathcal M})$ should be defined as a mapping from a
subset~$I$ and ${\mathcal M}$ to a nonnegative
number. However, according to the best of our knowledge,
no such quantities have ever been developed to define a workable
${\mathcal P}(I|{\mathcal M})$ for
an arbitrary subset $I$ in general parametric models, due to many
conceptual difficulties~\cite{Critchley-etal2001}.
Specifically, even though~\cite{Critchley-etal2001} placed the
Euclidean geometry on the perturbation space for one-sample problems,
such a geometrical structure cannot be easily generalizable
to general data structures (e.g., correlated data) and related
parametric models. For instance,
for correlated data, a sensible model
${\mathcal M}$ should model the correlation structure, and
a good measure ${\mathcal P}(I|{\mathcal M})$ should explicitly
incorporate the
correlation structure specified in ${\mathcal M}$ and the subset $I$.
However, the Euclidean geometry proposed by~\cite{Critchley-etal2001}
cannot incorporate the correlation structure in the correlated data.

Our choice of ${\mathcal P}(I|{\mathcal M})$ is motivated by five
principles, as follows:
\begin{itemize}
\item{(P.a)} (nonnegativity)
For any subset $I$, ${\mathcal P}(I|{\mathcal M})$ is always nonnegative.
\item{(P.b)} (uniqueness) ${\mathcal P}(I|{\mathcal M})=0$ if and only
if $I$ is an empty set.
\item{(P.c)} (monotonicity) If $I_2\subset I_1$, then ${\mathcal
P}(I_2|{\mathcal M})\leq{\mathcal P}(I_1|{\mathcal M})$.
\item{(P.d)} (additivity) If $I_2\subset I_1$, $ I_{1\cdot2}=I_1-I_2$ and
$p(\mathbf{Y}_{I_{1\cdot2}}| \mathbf{Y}_{[I_1]}, {\ve\theta})=p(\mathbf{
Y}_{I_{1\cdot2}}|\allowbreak \mathbf{Y}_{[I_{1\cdot2}]}, {\ve\theta})$ for all $\ve
\theta$, then we have
${\mathcal P}(I_1|{\mathcal M})={\mathcal P}(I_2|{\mathcal
M})+{\mathcal P}(I_{1\cdot2}|{\mathcal M})$.
\item{(P.e)}
${\mathcal P}(I|{\mathcal M})$ should naturally arise from the current
model ${\mathcal M}$, the data~$\mathbf{Y}$ and the subset $I$.
\end{itemize}
Principles (P.a) and (P.b) indicate that deleting any nonempty subset
always introduces a positive degree of perturbation.
Principle (P.c) indicates that deleting a larger subset always
introduces a larger degree of perturbation.
Principle (P.d) presents the condition for ensuring the additivity
property of the perturbation. Since $ \mathbf{Y}_{[I_{1\cdot2}]}$ is the
union of $\mathbf{Y}_{[I_1]}$ and $\mathbf{Y}_{I_2}$, $p(\mathbf{Y}_{I_{1\cdot
2}}| \mathbf{Y}_{[I_1]}, {\ve\theta})=p(\mathbf{Y}_{I_{1\cdot2}}| \mathbf{
Y}_{[I_{1\cdot2}]}, {\ve\theta})$ is equivalent to that of $\mathbf{
Y}_{I_{1\cdot2}}$ being independent of $\mathbf{Y}_{I_2}$ given $\mathbf{Y}_{[I_1]}$.
The additivity property has important implications in cross-sectional,
longitudinal and family data. For instance, in longitudinal data,
the degree of perturbation introduced by simultaneously deleting two
independent clusters equals the sum of
their degrees of individual cluster perturbations.

Principle (P.e) requires that ${\mathcal P}(I|{\mathcal M})$
depend on the triple $({\mathcal M}, \mathbf{Y}, I)$.
We propose ${\mathcal P}(I|{\mathcal M})$ based on the
Kullback--Leibler divergence between the fitted probability function
$p(\mathbf{Y}|{\ve\theta})$ and the probability function of a model for
characterizing the deletion of $\mathbf{Y}_I$, denoted by $p(\mathbf{Y}|{\ve
\theta}, I)$.
Note that $p(\mathbf{Y}|{\ve\theta})=p(\mathbf{Y}_{[I]}|{\ve\theta})p(\mathbf{
Y}_I|\mathbf{Y}_{[I]}, {\ve\theta})$, where $p(\mathbf{Y}_I|\mathbf{Y}_{[I]},
{\ve\theta})$ is the conditional density of $\mathbf{Y}_I$ given $\mathbf{Y}_{[I]}$.
Let ${\ve\theta}_* $ be the true value of $\ve\theta$ under ${\mathcal
M}$~\cite{White1982,White1994}.
We define $p(\mathbf{Y}|{\ve\theta}, I)$ as follows:
%
\begin{equation} \label{DPeq0}
p(\mathbf{Y}|{\ve\theta}, I)= p\bigl(\mathbf{Y}_{[I]}|{\ve\theta}\bigr)p\bigl(\mathbf{Y}_I|\mathbf{
Y}_{[I]}, {\ve\theta}_*\bigr),
\end{equation}
in which $p(\mathbf{Y}_I|\mathbf{Y}_{[I]}, {\ve\theta}_*)$ is independent of
${\ve\theta}$. In (\ref{DPeq0}), by fixing ${\ve\theta}={\ve\theta}_*$
in $p(\mathbf{Y}_I|\mathbf{Y}_{[I]}, {\ve\theta})$, we essentially drop the
information contained in $\mathbf{Y}_I$ as we estimate~$\ve\theta$. Specifically,
$ \hat{\ve\theta}_{[I]}$ is the maximum likelihood estimate of $\ve
\theta$ under $p(\mathbf{Y}|{\ve\theta}, I)$. If ${\mathcal M}$ is
correctly specified, then $p(\mathbf{Y}_I|\mathbf{Y}_{[I]}, {\ve\theta}_* )$
is the true data generator for $\mathbf{Y}_I$ given $\mathbf{Y}_{[I]}$.
The Kullback--Leibler distance between $p(\mathbf{Y}|{\ve\theta})$ and
$p(\mathbf{Y}|{\ve\theta}, I)$, denoted by $\operatorname{KL}(\mathbf{Y}, {\ve\theta
}|{\ve\theta}_*, I)$, is given by
%
\begin{equation} \label{DPeq1}
\quad\int p(\mathbf{Y}|{\ve\theta})\log\biggl(\frac{p(\mathbf{Y}|{\ve\theta
})}{p(\mathbf{Y}|{\ve\theta}, I)}\biggr)\,d\mathbf{Y}=\int p(\mathbf{Y}|{\ve\theta
})\log\biggl(\frac{p(\mathbf{Y}_I|\mathbf{Y}_{[I]}, {\ve\theta})}{p(\mathbf{
Y}_I|\mathbf{Y}_{[I]}, {\ve\theta}_* )}\biggr)\,d\mathbf{Y}.
\end{equation}
We use $ \operatorname{KL}(\mathbf{Y}, {\ve\theta}|{\ve\theta}_*, I)$ to measure
the effect of deleting $\mathbf{Y}_I$ on estimating~$\ve\theta$ without
knowing that the true value of $\ve\theta$ is $\ve\theta_*$.
If $\mathbf{Y}_I$ is independent of $\mathbf{Y}_{[I]}$, then we have
\[
\operatorname{KL}(\mathbf{Y}, {\ve\theta}|{\ve\theta}_*, I)=
\int p(\mathbf{Y}_I|{\ve\theta})\log\biggl(\frac{p(\mathbf{Y}_I| {\ve\theta
})}{p(\mathbf{Y}_I| {\ve\theta}_* )}\biggr)\,d\mathbf{Y}_I,
\]
which is independent of $\mathbf{Y}_{[I]}$. In this case, the effect of
deleting $\mathbf{Y}_I$ on estimating~$\ve\theta$ only depends on $\{
p(\mathbf{Y}_I|{\ve\theta})\dvtx  {\ve\theta} \in\Theta\}$.

A conceptual difficulty associated with $ \operatorname{KL}(\mathbf{Y}, {\ve\theta
}|{\ve\theta}_*, I)$ is that
both
${\ve\theta}$ and~${\ve\theta}_*$ are unknown.
Although ${\ve\theta}_*$ is unknown, it can be assumed to be a fixed
value from a frequentist viewpoint.
For the unknown $\ve\theta$, we can always use the data $\mathbf{Y}$ and
the current model ${\mathcal M}$ to calculate
an estimator $\hat{\ve\theta}$ in a neighborhood of ${\ve\theta}_*$.
Under some mild conditions~\cite{White1982,White1994},
one can show that $\sqrt{n}(\hat{\ve\theta}-{\ve\theta}_* )$ is
asymptotically normal, and thus $\hat{\ve\theta}$ should be
centered around $\ve\theta_*$. Moreover, since Cook's distance is to
quantify the change of the parameter estimates after deleting a subset, we
need to consider all possible~$\ve\theta$ around $\ve\theta_*$, instead
of focusing on a single
$\ve\theta$.
Specifically, we consider~$\ve\theta$ in a neighborhood of~${\ve\theta
}_* $ by assuming a Gaussian prior for $\ve\theta$ with mean~${\ve\theta
}_* $ and positive definite covariance matrix $\Sigma_{*}$ (e.g., the
Fisher information matrix), denoted by $p({\ve\theta}| {\ve\theta}_* ,
\Sigma_{*})$.
Finally, we define ${\mathcal P}(I|{\mathcal M})$ as the weighted
Kullback--Leibler distance between $p(\mathbf{Y}|{\ve\theta})$ and $p(\mathbf{
Y}|{\ve\theta}, I)$
as follows:
%
\begin{equation}\label{DPeq2}
{\mathcal P}(I|{\mathcal M})=\int \operatorname{KL}(\mathbf{Y}, {\ve\theta}|{\ve
\theta}_*, I)p({\ve\theta}| {\ve\theta}_* , \Sigma_*) \,d\ve\theta.
\end{equation}
This quantity ${\mathcal P}(I|{\mathcal M})$ can also be interpreted as
the average effect of deleting $\mathbf{Y}_I$ on estimating $\ve\theta$
with the prior information that the estimate of $\ve\theta$ is centered
around $\ve\theta_*$.
Since ${\mathcal P}(I|{\mathcal M})$ is directly calculated from the
model ${\mathcal M}$ and the subset~$I$,
it can naturally account for any structure specified in ${\mathcal M}$.
Furthermore, if we are interested in a particular set of components of
$\ve\theta$ and treat others as
nuisance parameters, we may fix these nuisance parameters at their true value.

To compute $ {\mathcal P}(I|{\mathcal M})$ in a real data analysis, we
only need to specify~${\mathcal M}$ and $({\ve\theta}_* , \Sigma_*)$.
Then we may use
some numerical integration methods to compute ${\mathcal
P}(I|{\mathcal M})$. Although $({\ve\theta}_* , \Sigma_*)$ are unknown,
we suggest substituting~${\ve\theta}_*$ by an estimator of~$\ve\theta
$, denoted by~$\tilde{\ve\theta}$, and~$\Sigma_{*}$ by the covariance matrix of~$\tilde{\ve\theta}$.
Throughout the paper,
since $\hat{\ve\theta}$ is a consistent estimator of $\ve\theta_*$
\cite{White1982,White1994},
we set $\tilde{\ve\theta}=\hat{\ve\theta}$ and $\Sigma_*$ as the
covariance matrix of $\hat{\ve\theta}$.


We obtain the following theorems, whose detailed assumptions and proofs
can be found in the \hyperref[app]{Appendix}.

\begin{theorem}\label{th1} Suppose that $L(\{\mathbf{Y}\dvtx  p(\mathbf{
Y}_I|\mathbf{Y}_{[I]}, {\ve\theta})=p(\mathbf{Y}_I|\mathbf{Y}_{[I]}, {\ve\theta
}_* )\})>0$ for any
$\ve\theta\not={\ve\theta}_* $, where $L(A)$ is the Lebesgue measure of
a set $A$. Then, ${\mathcal P}(I|{\mathcal M})$ defined in (\ref{DPeq2})
satisfies the four principles \textup{(P.a)--(P.d)}.
\end{theorem}


As an illustration, we show how to calculate ${\mathcal P}(I|{\mathcal
M})$ under the standard linear regression model for cross-sectional
data as follows.

\begin{example}\label{ex1}
$\!\!\!$Consider the linear regression model
$y_i=\mathbf{x
}_i^T{\ve\beta}_*+\varepsilon_i$, where~$\mathbf{x}_i$ is a $p\times1$
vector, and the
$\varepsilon_i$ are independently and identically distributed (i.i.d.) as
$N(0, \sigma_*^2)$.
Let $\mathbf{y}=(y_1, \ldots, y_n)^T$ and $\mathbf{X}$
be an $n\times p$ matrix of rank $p$ with $i$th row $\mathbf{x}_i^T$.
In this case, $\ve\theta=({\ve\beta}^T, \sigma^2)^T$.
Recall that
$\hat{\ve\beta}=(\mathbf{X}^T\mathbf{X})^{-1}\mathbf{X}^T\mathbf{y}$, $\hat\sigma
^2=\mathbf{y}^T(\mathbf{I}_n-H_x)\mathbf{y}/n$,
$\operatorname{Cov}(\hat{\ve\beta})=\sigma_*^2(\mathbf{X}^T\mathbf{X})^{-1}$ and
$\operatorname{var}(\hat\sigma^2)=2\sigma_*^4/n$, where~$\mathbf{I}_n$ is an $n\times n$ identity matrix and $H_x=(h_{ij})=\mathbf{
X}(\mathbf{X}^T\mathbf{X})^{-1}\mathbf{X}^T$.
We first compute the degree of the perturbation for
deleting each $(y_i,
\mathbf{x}_i)$. We consider two scenarios: fixed and random covariates.
For the case of fixed covariates,~${\mathcal M}$ assumes\vadjust{\goodbreak} $y_i\sim
N(\mathbf{x}_i^T{\ve\beta}, \sigma^2)$. After some algebraic calculations,
it can be shown that $ {\mathcal P}(\{i\}|{\mathcal M})$ equals
%
\begin{equation}
\quad 0.5E_\theta[\log(\sigma_*^2/\sigma^2)]+0.5
\frac{ \mathbf{x}_i^TE_\theta[({\ve\beta}-{\ve\beta}_*)({\ve\beta}-{\ve
\beta}_*)^T]\mathbf{x}_i }{\sigma_*^2}
\approx\frac1{2n} + \frac12 h_{\mathit{ii}}, \label{Exeq1}
\end{equation}
where $E_\theta$ is taken with respect to $p({\ve\theta}| {\ve\theta}_*
, G_{n\theta}^{-1})$.
Moreover, the right-hand side of~(\ref{Exeq1}) contains only terms
involving $n$ and $\mathbf{X}$, since perturbation is defined only in
terms of the underlying model ${\mathcal M}$. This is also at
the core of why only stochastic ordering is possible for Cook's distance,
which is a~function of both the perturbation and the data.
See Section~\ref{sec2.3} for detailed discussions.
Furthermore, if ${\ve\beta}$ is the parameter of interest in ${\ve\theta
}$ and
$\sigma^2$ is a nuisance parameter, then $0.5E_\theta[\log(\sigma
_*^2/\sigma^2)]$, and $1/(2n)$ can be dropped
from $ {\mathcal P}(\{i\}|{\mathcal M})$ in (\ref{Exeq1}).

Furthermore, for the case of random covariates, we assume that the
$\mathbf{x}_i$'s are independently and identically distributed with mean
$\mu_x$ and covariance
matrix $\Sigma_x$. It can be shown that $ {\mathcal P}(\{i\}|{\mathcal
M})$ equals
%
\begin{equation}
 0.5E_\theta[\log(\sigma_*^2/\sigma^2)]+0.5 \sigma_*^{-2}{ \operatorname{tr}\{
\Sigma_x E_\theta[({\ve\beta}-{\ve\beta}_*)({\ve\beta}-{\ve\beta
}_*)^T]\} }\approx \frac{1}{2n}+\frac{p}{2n}. \label{Exeq2}\hspace*{-35pt}
\end{equation}
If $\ve\beta$ is the parameter of interest in $\ve\theta$, and
$\sigma^2$ is a nuisance parameter, then $ {\mathcal P}(\{i\}|{\mathcal
M})$ reduces to $p/(2n)$.
Furthermore, consider deleting a subset of observations $\{ (y_{i_k},
\mathbf{x}_{i_k})\dvtx  k=1, \ldots, n(I)\}$ and $I=\{i_1, \ldots, i_{n(I)}\}$.
It follows from Theorem~\ref{th1} that
${\mathcal P}(\{i_1, \ldots, i_{n(I)}\}|{\mathcal M})=\sum
_{k=1}^{n(I)} {\mathcal P}(\{i_k\}|{\mathcal M})$.
Furthermore, for the case of random covariates, we have
$
{\mathcal P}(I|{\mathcal M})=n(I) {\mathcal P}(\{1\}|{\mathcal M})
$
for any subset $I$ with $n(I)$ observations.
Thus, in this case, deleting any two subsets~$I_1$ and~$I_2$ with the
same number of observations, that is, $n(I_1)= n(I_2)$, has the same
degree of perturbation.
An important implication of these calculations in real data analysis is
that we can directly compare $\operatorname{CD}({I}_1)$ and $\operatorname{CD}({ I}_2)$
when $n(I_1)= n(I_2)$.
\end{example}

\subsection{Cook's distance and degree of perturbation}\label{sec2.3}

To understand
the relationship between
${\mathcal P}(I|{\mathcal M})$ and $\operatorname{CD}({ I})$ in (\ref{size1}),
we temporarily assume that the fitted model ${\mathcal M}$ is the true
data generator of $\mathbf{Y}$.
To have a better understanding of Cook's distance, we consider the
standard linear regression model for cross-sectional data as follows.

\setcounter{example}{0}
\begin{example}[(Continued)]\label{ex1c}
We are interested in ${\ve\beta
}$ and treat $\sigma^2$ as a nuisance parameter.
We first consider deleting individual observations
in linear regression. Cook's distance~\cite{Cook1977} for case
$i$, $(y_i, \mathbf{x}_i)$, is given by
%
\begin{equation}\label{chap5eq9}
\operatorname{CD}(\{i\})=\frac{(\hat{{\ve\beta}}-\hat{{\ve\beta}}_{[i]})^T \mathbf{
X}^T\mathbf{X}
(\hat{{\ve\beta}}-\hat{{\ve\beta}}_{[i]})} { \hat\sigma^2}=\frac{\sigma
^2}{ \hat\sigma^2}
t_i^2\frac{h_{\mathit{ii}}}{1-h_{\mathit{ii}}},
\end{equation}
where $\hat{\ve\beta}$ is the least squares estimate of ${\ve\beta}$,
$\hat\sigma^2$ is a consistent estimator of $\sigma^2$, $t_i=
{\hat e_i}/(\sigma\sqrt{1-h_{\mathit{ii}}})$ and $ \hat{\ve\beta}_{[i]}=\hat{\ve
\beta}-
{(\mathbf{X}^T\mathbf{X})^{-1}\mathbf{x}_i\hat e_i}/(1-h_{\mathit{ii}})$, in which
$\hat e_i=y_i-\mathbf{x}_i^T\hat{\ve\beta}$. It should be noted that
except for a constant
$p$, $\operatorname{CD}(\{i\})$ is almost the same as the original Cook's
distance (Cook~\cite{Cook1977}).
As shown in~(\ref{Exeq1}) and~(\ref{Exeq2}), regardless of the exact
value of $(y_i,
\mathbf{x}_i)$, deleting any $(y_i,
\mathbf{x}_i)$ has approximately the same degree of perturbation to
${\mathcal M}$.
Moreover, the
$\operatorname{CD}(\{i\})$ are comparable regardless of~$i$.
Specifically, if $\varepsilon_i\sim N(0, \sigma^2)$, then~$t_i^2$
follows the $\chi^2(1)$ distribution for all $i$. For the case of
random covariates, if~$\mathbf{x}_i$ are identically distributed, then all
$\operatorname{CD}(\{i\})$ are truly
comparable, since they follow the same distribution.

We consider deleting multiple
observations in the linear model. Cook's distance
for deleting the subset $I$ with $n(I)$ is given by
%
\begin{equation}\label{chap5eq10}
\frac{(\hat{{\ve\beta}}-\hat{{\ve\beta}}_{[I]})^T \mathbf{X}^T\mathbf{X}
(\hat{{\ve\beta}}-\hat{{\ve\beta}}_{[I]})}
{ \hat\sigma^2}=\frac{1}{ \hat\sigma^2}
\hat\mathbf{e}_I^T\bigl(\mathbf{I}_{n(I)}-H_I\bigr)^{-1}H_I\bigl(\mathbf{
I}_{n(I)}-H_I\bigr)^{-1}\hat\mathbf{e}_I,\hspace*{-45pt}
\end{equation}
where $\hat\mathbf{e}_I$ is an ${n(I)}\times1$ vector containing all $\hat
e_i$ for $i\in I$
and $H_I=\break\mathbf{X}_I(\mathbf{X}^T\mathbf{X})^{-1}\mathbf{X}_I^T$, in which
$\mathbf{X}_I$ is an ${n(I)}\times p$ matrix whose rows are $\mathbf{x}_i^T$
for all $i\in I$. Similar to the deletion of a single case, deleting
any subset with the same number of observations introduces
approximately the same degree of perturbation to~${\mathcal M}$,
and the
$\operatorname{CD}(I)$ are comparable among all subsets with the same $n(I)$.
We will make this statement precise in Theorem~\ref{th2} given below.

Generally, we want to compare $\operatorname{CD}(I_1)$ and $\operatorname{CD}(I_2)$ for
any two subsets with $n(I_1)\not= n(I_2)$.
As shown in Example~\ref{ex1}, when $n(I_1)>n(I_2)$, deleting~$I_1$ introduces
a larger degree of perturbation to model ${\mathcal M}$ compared to
deleting~$I_2$.
To compare Cook's distances among arbitrary subsets, we need to
understand the relationship
between ${\mathcal P}(I|{\mathcal M}) $ and $\operatorname{CD}(I)$ for any
subset~$I$.
Surprisingly, in linear regression for cross-sectional data, we can
show the
stochastic relationship between ${\mathcal P}(I|{\mathcal M}) $ and
$\operatorname{CD}(I)$, as
follows.
\end{example}

\begin{theorem}\label{th2} For the standard linear model, where
$\mathbf{y}=\mathbf{
X}{\ve\beta}+\varepsilon$ and $\varepsilon\sim N(\mathbf{0}, \sigma^2I_n)$, we
have the following results:

\begin{longlist}[(a)]
\item[(a)]
For any $I_2\subset I_1$,
$\operatorname{CD}(I_1)$ is stochastically larger than $\operatorname{CD}(I_2)$ for
any~$\mathbf{X}$, that is, $
\mathrm{P}(\operatorname{CD}(I_1)>t|{\mathcal M})\geq
\mathrm{P}(\operatorname{CD}(I_2)>t|{\mathcal M})$ holds for any $t\geq0$.

\item[(b)] Suppose that the components of $\mathbf{X}_I$ and $\mathbf{X}_{I'}$ are
identically distributed for any two subsets $I$ and $I'$ with $n(I)=n(I')$.
Thus, $\operatorname{CD}(I)$ and $\operatorname{CD}(I')$ follow the same distribution
when $n(I)=n(I')$ and
$\operatorname{CD}(I_1)$ is stochastically larger than $\operatorname{CD}(I_2)$ for
any two subsets $I_2$ and
$I_1$ with $n(I_1)>n(I_2)$.
\end{longlist}
\end{theorem}

Theorem~\ref{th2}(a)
shows that the Cook's distances for two nested subsets
satisfy the stochastic ordering property.
Theorem~\ref{th2}(b) indicates that for random covariates, the Cook's
distances for any
two subsets also satisfy the stochastic ordering property under some
mild conditions.

According to Theorem~\ref{th2}, for more complex data structures and models, it
may be natural to use the stochastic order
to stochastically quantify the positive\vadjust{\goodbreak}
association between the degree of the perturbation and the magnitude of
Cook's distance. Specifically,
we consider two possibly overlapping subsets ${I}_1$ and ${ I}_2$
with ${\mathcal P}(I_1|{\mathcal M}) > {\mathcal P}(I_2|{\mathcal
M})$. Although
$\operatorname{CD}(I_1)$ may not be greater than $\operatorname{CD}(I_2)$ for a fixed
dataset $\mathbf{Y}$,
$\operatorname{CD}(I_1)$, as a random variable, should be \textit{stochastically
larger} than $\operatorname{CD}(I_2)$ if ${\mathcal M}$ is the true model for
$\mathbf{Y}$. We make the following assumption.

\renewcommand{\theassumption}{A\arabic{assumption}}
\begin{assumption}\label{assa1}
$\!\!\!$For any two subsets $I_1$ and $I_2$
with ${\mathcal P}(I_1|{\mathcal M}) > {\mathcal P}(I_2|{\mathcal
M})$,
%
\begin{equation}\label{size2}
\mathrm{P}\bigl(\operatorname{CD}(I_1)>t|{\mathcal M}\bigr)  \geq
\mathrm{P}\bigl(\operatorname{CD}(I_2)>t|{\mathcal M}\bigr)
\end{equation}
holds for any $t>0$, where the probability is taken with respect to
${\mathcal M}$.
\end{assumption}

Assumption~\ref{assa1} is essentially saying that if
${\mathcal M}$ is the true data generator, then $\operatorname{CD}(I_1)$
stochastically dominates $\operatorname{CD}(I_2)$ whenever
${\mathcal P}(I_1|{\mathcal M})>{\mathcal P}(I_2|{\mathcal M})$.
According to the definition of stochastic ordering~\cite{Shaked06}, we
can now obtain the following proposition.

\begin{proposition}\label{prop1} Under Assumption~\ref{assa1}, for any two
subsets $I_1$ and $I_2$ with
${\mathcal P}(I_1|{\mathcal M})>{\mathcal P}(I_2|{\mathcal M})$, Cook's
distance satisfies
%
\begin{equation}\label{size3}
E[h(\operatorname{CD}(I_1))|{\mathcal M}]   \geq   E[h(\operatorname
{CD}(I_2))|{\mathcal
M}]
\end{equation}
and holds for all increasing functions $h(\cdot)$. In particular, we
have $ E[ \operatorname{CD}(I_1) |\allowbreak{\mathcal M}]\geq E[ \operatorname{CD}(I_2)
|{\mathcal M}] $ and $Q_{\operatorname{CD}(I_1)}(\alpha|{\mathcal M})$ is
greater than the $\alpha$-quantile of
$Q_{\operatorname{CD}(I_2)}(\alpha|{\mathcal M})$ for any $\alpha\in[0,
1]$, where $Q_{\operatorname{CD}(I)}(\alpha|{\mathcal M})$ denotes the
$\alpha$-quantile of the distribution of $\operatorname{CD}(I)$ for any
subset $I$.
\end{proposition}

Proposition~\ref{prop1} formally characterizes the fundamental issue of Cook's
distance. Specifically, for
any two subsets ${I}_1$ and ${ I}_2$ with
${\mathcal P}(I_1|{\mathcal M})>{\mathcal P}(I_2|{\mathcal M})$,
$\operatorname{CD}({ I}_1)$ has a high probability of being greater than
$\operatorname{CD}({ I}_2)$ when ${\mathcal M}$ is the true data generator.
Thus, Cook's distance for
subsets with different degrees of perturbation are not directly comparable.
More importantly, it indicates that
$\operatorname{CD}({ I})$ cannot be simply expressed as a linear function of
${\mathcal P}(I|{\mathcal M})$.
Thus,
the standard solution, which standardizes $\operatorname{CD}({ I})$ by
calculating the ratio of $\operatorname{CD}({ I})$ over ${\mathcal
P}(I|{\mathcal M})$,
is not desirable for controlling for the effect of ${\mathcal
P}(I|{\mathcal M})$.

\subsection{Scaled Cook's distances}\label{sec2.4}

We focus on developing several scaled\break Cook's distances for $I$,
denoted by $\operatorname{SCD}(I)$, to
detect relatively influential subsets, while accounting for the degree
of perturbation ${\mathcal P}(I|{\mathcal M})$.
Since we have characterized the stochastic relationship between
${\mathcal P}(I|{\mathcal M})$ and $\operatorname{CD}({ I})$ when~${\mathcal M}$
is the true data generator, we are interested in reducing the effect of
the difference among ${\mathcal P}(I|{\mathcal M})$ for different
subsets $I$ on the magnitude of $\operatorname{CD}({ I})$.
A simple solution is to calculate several features (e.g., mean, median,
or quantiles)
of $\operatorname{CD}(I)$ and match them across different subsets under the
assumption that
${\mathcal M}$ is the true data generator.
Throughout the paper, we consider\vadjust{\goodbreak}
two pairs of features
including (mean, Std) and (median, Mstd), where Std and Mstd,
respectively, denote the standard deviation and the median
standard deviation.
By matching any of the two pairs, we can
at least ensure that the centers and scales of the scaled Cook's
distances for
different subsets are the same when ${\mathcal M}$ is the true data generator.

We introduce two scaled Cook's distance measures, called scaled Cook's
distances, as follows.

\begin{definition}\label{def1}
The \textit{scaled Cook's distances} for
matching (mean, Std) and (median, Mstd) are, respectively, defined as
\begin{eqnarray*}
\operatorname{SCD}_1(I)&=&\frac{\operatorname{CD}(I)-E[\operatorname{CD}(I)|{\mathcal M}]}{ {\operatorname
{Std}[\operatorname{CD}(I)|{\mathcal M}]}}
\quad\mbox{and}\\
 \operatorname{SCD}_2(I)&=&\frac{\operatorname{CD}(I)-Q_{\operatorname{CD}(I)}(0.5|{\mathcal
M})}{ {\operatorname{Mstd}[\operatorname{CD}(I)|{\mathcal M}]}},
\end{eqnarray*}
where both the expectation and the quantile are taken with respect to
${\mathcal M}$.
\end{definition}

We can use $\operatorname{SCD}_1(I)$ and $\operatorname{SCD}_2(I)$ to evaluate
the relatively influential level for different subsets $I$. A large
value of
$\operatorname{SCD}_1(I)$ [or $\operatorname{SCD}_2(I)$] indicates that the subset $I$
is relatively influential.
Therefore, for any two subsets $I_1$
and $I_2$, the probability of observing the event
$\operatorname{SCD}(I_1)>\operatorname{SCD}(I_2)$ and that of the event
$\operatorname{SCD}(I_1)<\operatorname{SCD}(I_2)$ should be reasonably close to each
other. Thus, the $\operatorname{SCD}(I)$ are roughly comparable.
Note that the scaled Cook's distances do not
provide a ``per unit'' effect of removing one observation within the
set $I$, whereas
they measure the standardized influential level of the set $I$ when
$\mathcal M$ is true.
Moreover, the standardization in Definition~\ref{def1} still implies that higher
than average values of $\operatorname{CD}(I)$ still correspond with high
positive values
of $\operatorname{SCD}(I)$, even though for some deletions, it is possible for
$\operatorname{SCD}(I)$ to be negative unlike $\operatorname{CD}(I)$.

The next task is how to compute $E[\operatorname{CD}(I)|{\mathcal M}]$, ${\operatorname
{Std}[\operatorname{CD}(I)|{\mathcal M}]}$,\break
$\operatorname{Mstd}[\operatorname{CD}(I)| {\mathcal M}]$ and
$Q_{\operatorname{CD}(I)}(0.5|{\mathcal M})$ for each subset $I$ under the assumption
that ${\mathcal M}$ is the true data generator.
Computationally, we suggest using the parametric bootstrap to
approximate the four quantities of $\operatorname{CD}(I)$ as follows:

Step 1. We use $\hat{\mathcal M}=\{p(\mathbf{Y}|\hat{\ve\theta})\}$ to
approximate the model ${\mathcal M}=\{p(\mathbf{Y}|{\ve\theta}_* )\}$,
generate a random sample $\mathbf{Y}^s$ from $p(\mathbf{Y}|\hat{\ve\theta})$
and then calculate $\operatorname{CD}(I)^{(s)}=F_1(I, \hat{\mathcal M}, \mathbf{Y}^s)$
for each $s$ and each subset $I$.

Step 2. By repeating Step 1 $S$ times, we can obtain
a sample $\{\operatorname{CD}(I)^{(s)}\dvtx  s=1, \ldots, S\}$ and then we use its
empirical mean $\overline{\operatorname{CD}(I)}=\sum_{s=1}^S\operatorname
{CD}(I)^{(s)}/S$ to approximate $E[\operatorname{CD}(I)|{\mathcal
M}]$.

Step 3. We approximate ${\operatorname{Std}[\operatorname{CD}(I)|{\mathcal M}]}$,
$Q_{\operatorname{CD}(I)}(0.5|{\mathcal M})$ and
${\operatorname{Mstd}[\operatorname{CD}(I)|{\mathcal M}]}$ by using their corresponding empirical quantities of $\{\operatorname
{CD}(I)^{(s)}\dvtx  s=1, \ldots, S\}$.

In this process, we have used
$\hat{\mathcal M} $ to approximate ${\mathcal M}$~\cite{White1982} and simulated data~$\mathbf{Y}^s$ from
$\hat{\mathcal M}$\vadjust{\goodbreak}
in the standard parametric bootstrap method. If $\mathbf{Y}$ truly comes
from ${\mathcal M}$, then
the simulated data $\mathbf{Y}^s$ should resemble $\mathbf{Y}$.
Since $\hat{\ve\theta}$ is a consistent estimate of ${\ve\theta}_* $,
$ E[F_1(I, \hat{\mathcal M}, \mathbf{Y})| \hat{\mathcal M}]\approx
E[F_1(I, {\mathcal M}, \mathbf{Y})| {\mathcal M}]$
and thus $\overline{\operatorname{CD}(I)}$ is a consistent estimate of
$E[F_1(I, {\mathcal M}, \mathbf{Y})| {\mathcal M}]$. Similar arguments
hold for the other
three quantities of $ {\operatorname{CD}(I)}$.
In Steps 2 and 3, we can use a moderate $S$, say $S=100$, in order to
accurately approximate all four quantities of $ {\operatorname{CD}(I)}$.
According to our experience, such an approximation is very accurate,
even for small $n$. See the simulation studies in Section~\ref{sec3.1} for details.
However, for most statistical models with
complex data structures, it can be computationally intensive
to compute $\hat{\ve\theta}^s$ for each $\mathbf{Y}^s$. We will address
this issue in Section~\ref{sec2.6}.

As an illustration, we consider how to calculate $\operatorname{SCD}_1(I)$ for
any subset~$I$
in the linear regression model.

\setcounter{example}{0}
\begin{example}[(Continued)]
In (\ref{chap5eq10}), since all
${\operatorname{CD}}(I)$ share $\hat\sigma^2$, we replace~$\hat\sigma^2$ by
$\sigma_*^2$. Thus, we approximate ${\operatorname{CD}}(I)$ by
$\operatorname{CD}_*(I)={\ve\varepsilon}^T\mathbf{W}_{*}{\ve\varepsilon}/\sigma_*^2$,
where ${\ve\varepsilon}=(\varepsilon_{1}, \ldots, \varepsilon_{n})^T\sim N(\mathbf{
0}, \sigma_*^2\mathbf{I}_n)$ and
\[
\mathbf{W}_{*} = (\mathbf{I}_n- H_{x})U_I\bigl(\mathbf{I}_{n(I)}-
H_{I}\bigr)^{-1}H_{I}\bigl(\mathbf{I}_{n(I)}-H_{I}\bigr)^{-1}U_I^T
(\mathbf{I}_n- H_{x}).
\]
To compute $\operatorname{SCD}_1(I)$, we just need to calculate
the two quantities $E[\operatorname{CD}_*(I)|{\mathcal M}]$ and ${\operatorname
{Std}[\operatorname{CD}_*(I)|{\mathcal M}]}$.
Since $\operatorname{CD}_*(I)$ is a quadratic form, it can be shown that
\begin{eqnarray*}
 E[ \operatorname{CD}_*(I)|{\mathcal M}] &=& E\bigl\{\operatorname{tr}\bigl[\bigl(\mathbf{
I}_{n(I)}-H_{I}\bigr)^{-1}\bigr]\big|{\mathcal M}_X\bigr\}-{n(I)}, \nonumber
\\
 \operatorname{Var}[ \operatorname{CD}_*(I)|{\mathcal M}]&=&
\operatorname{Var}\bigl\{\operatorname{tr}\bigl[\bigl(\mathbf{I}_{n(I)}-H_{I}\bigr)^{-1}\bigr]\big|{\mathcal M}_X\bigr\}
\\
&&{}+2E\bigl\{\operatorname{tr}\bigl[\bigl\{\bigl(\mathbf{
I}_{n(I)}-H_{I}\bigr)^{-1}H_I\bigr\}^2\bigr]\big|{\mathcal M}_X\bigr\}, \nonumber
\end{eqnarray*}
where $E[\cdot|{\mathcal M}_X]$ denotes the expectation taken with
respect to $\mathbf{X}$.
\end{example}
\subsection{Conditionally scaled Cook's distances}\label{sec2.5}

In certain research settings (e.g., regression), it may be better to
perform influence analysis
while fixing some covariates of interest, such
as measurement time. For instance, in longitudinal data,
if different subjects can have different numbers of measurements and
measurement times, which are not covariates of interest in an
influence analysis, it may be better to eliminate their effect
in calculating Cook's distance.
We are interested in comparing Cook's distance relative to ${\mathcal
P}(I|{\mathcal M})$ while fixing some covariates.

To eliminate the effect of some fixed covariates,
we introduce two conditionally scaled Cook's distances as follows.

\begin{definition}\label{def2} The \textit{conditionally scaled Cook's
distances} (CSCD) for matching (mean, Std) and (median, Mstd) while
controlling for $\mathbf{Z}$ are,
respectively, defined as
\begin{eqnarray*}
\operatorname{CSCD}_1(I, \mathbf{Z})&=&\frac{\operatorname{CD}(I)-E[\operatorname
{CD}(I)|{\mathcal M}, \mathbf{Z}]}{\operatorname{Std}[\operatorname{CD}(I)|{\mathcal M},
\mathbf{Z}]},\vadjust{\goodbreak} \\
\operatorname{CSCD}_2(I, \mathbf{Z})&=&\frac{\operatorname{CD}(I)-\operatorname
{Q}_{\operatorname{CD}(I)}(0.5|{\mathcal M}, \mathbf{Z})}{\operatorname{Mstd}[\operatorname
{CD}(I)|{\mathcal M}, \mathbf{Z}]},
\end{eqnarray*}
where $\mathbf{Z}$ is the set of some fixed covariates in $\mathbf{Y}$ and the
expectation and quantiles are taken with respect to
${\mathcal M}$ given $\mathbf{Z}$.\vspace*{-3pt}
\end{definition}

According to Definition~\ref{def2}, these
conditionally scaled Cook's distances can be used to evaluate the relative
influential level of different subsets~$I$ given $\mathbf{Z}$. Similar to
$\operatorname{SCD}_1(I)$ and $\operatorname{SCD}_2(I)$, a large value of
$\operatorname{CSCD}_1(I, \mathbf{Z})$ [or $\operatorname{CSCD}_2(I, \mathbf{Z})$]
indicates a large influence of the subset $I$ after controlling for~$\mathbf{Z}$. It should be noted that because $\mathbf{Z}$ is fixed,
the $\operatorname{CSCD}_k(I, \mathbf{Z})$ do not reflect the influential level of~$\mathbf{
Z}$, and the $\operatorname{CSCD}_k(I, \mathbf{Z})$ may vary across different~$\mathbf{
Z}$. The conditionally scaled Cook's distances measure the difference
of the observed influence level of the set~$I$
given~$\mathbf{Z}$ to the expected influence level of a set with
the same data structure when~$\mathcal M$ is true and $\mathbf{Z}$ is fixed.

The next problem is how to compute $E[\operatorname{CD}(I)|{\mathcal M}, \mathbf{
Z}]$, ${\operatorname{Std}[\operatorname{CD}(I)|{\mathcal M}, \mathbf{Z}]}$,
$Q_{\operatorname{CD}(I)}(0.5|{\mathcal M}, \mathbf{Z})$ and ${\operatorname{Mstd}[\operatorname
{CD}(I)|{\mathcal M}, \mathbf{Z}]}$ for each subset $I$ when ${\mathcal
M}$ is the true data generator and $\mathbf{Z}$ is fixed. Similar to the
computation of the scaled Cook's distances, we can essentially use
almost the same approach to
approximate the four quantities for $\operatorname{CSCD}_1(I, \mathbf{Z})$ and
$\operatorname{CSCD}_2(I, \mathbf{Z})$. However, a slight difference occurs in
the way that we simulate the data. Specifically, let $\mathbf{Y}_Z$ be the
data $\mathbf{Y}$ with~$\mathbf{Z}$ fixed. We need to simulate random samples
$\mathbf{Y}_Z^s$ from
$ \hat{\mathcal M}_Z=\{p(\mathbf{Y}_Z|\mathbf{Z}, \hat{\ve\theta})\}$ and
then calculate $\operatorname{CD}(I)^{(s)}=F_1(I, \hat{\mathcal M}_Z, (\mathbf{Y}_Z^s,
\mathbf{Z}))$ for each subset $I$.

As an illustration, we consider how to calculate $\operatorname{CSCD}_1(I, \mathbf{
Z})$ for any subset $I$
in the linear regression model.\vspace*{-3pt}

\setcounter{example}{0}
\begin{example}[(Continued)]
We set $\mathbf{Z}=\mathbf{X}$ to calculate $\operatorname{CSCD}_1(I, \mathbf{Z})$. We
need to compute $E[\operatorname{CD}_*(I)|{\mathcal M}, \mathbf{Z}]$ and
${\operatorname{Std}[\operatorname{CD}_*(I)|{\mathcal M}, \mathbf{Z}]}$. Since $\operatorname
{CD}_*(I)$ is a~quadratic form, it is easy to show
$E[ \operatorname{CD}_*(I)|{\mathcal M}] = \operatorname{tr}[(\mathbf{
I}_{n(I)}-H_{I})^{-1}]-{n(I)}$ and
$ \operatorname{Var}[ \operatorname{CD}_*(I)|{\mathcal M}]=
2 \operatorname{tr}[\{(\mathbf{
I}_{n(I)}-H_{I})^{-1}H_I\}^2]. $\vspace*{-3pt}
\end{example}

\subsection{First-order approximations}\label{sec2.6}

We have focused on developing the\break scaled Cook's distances and their
approximations for the
linear regression model.
More generally, we are interested in approximating the scaled Cook's
distances for a large class of parametric models for both independent
and dependent data.

We obtain the following theorem.\vspace*{-3pt}

\begin{theorem}\label{th3}If Assumptions~\ref{assa2}--\ref{assa5} in the
\hyperref[app]{Appendix} hold and $n(I)/n\rightarrow\gamma\in[0, 1)$, where
$n(I)$ denotes the number of observations of $I$, then we have the following
results:
\begin{longlist}[(a)]
\item[(a)] Let $\mathbf{F}_{n}({\ve\theta})=-\partial_\theta^2\log
p(\mathbf{Y}|
{\ve\theta})$, $ \mathbf{f}_{I}({\ve\theta})=\partial_\theta\log
p(\mathbf{Y}_I|\mathbf{Y}_{[I]}, \hat{\ve\theta})$ and $\mathbf{s}_{I}({\ve\theta
})=-\partial_\theta^2 \log p(\mathbf{
Y}_I|\mathbf{Y}_{[I]}, {\ve\theta})$, $ \operatorname{CD}(I)$ can be approximated by
%
\begin{equation}\qquad
\widetilde{\operatorname{CD}}(I) = \mathbf{f}_{I}(\hat{\ve\theta})^T[\mathbf{
F}_{n}(\hat{\ve\theta})-\mathbf{s}_{I}(\hat{\ve\theta})]^{-1}
\mathbf{F}_{n}(\hat{\ve\theta})
[\mathbf{F}_{n}(\hat{\ve\theta})-\mathbf{s}_{I}(\hat{\ve\theta})]^{-1} \mathbf{
f}_{I}(\hat{\ve\theta});\vadjust{\goodbreak}
\end{equation}

\item[(b)]
$E[\widetilde{\operatorname{CD}}(I)|{\mathcal M}]\approx
\operatorname{tr} (\{E[\mathbf{F}_{n}(\hat{\ve\theta})|{\mathcal M}]-E[\mathbf{
s}_{I}(\hat{\ve\theta})|{\mathcal M}]\}^{-1}E[\mathbf{
s}_{I}(\hat{\ve\theta})|{\mathcal M}]); $

\item[(c)]
$E[\widetilde{\operatorname{CD}}(I)|{\mathcal M}, \mathbf{Z}]\approx
\operatorname{tr} (\{E[\mathbf{F}_{n}(\hat{\ve\theta})|{\mathcal M}, \mathbf{
Z}]-E[\mathbf{
s}_{I}(\hat{\ve\theta})|{\mathcal M}, \mathbf{Z}]\}^{-1}E[\mathbf{
s}_{I}(\hat{\ve\theta})|\break {\mathcal M}, \mathbf{Z}]).$
\end{longlist}
\end{theorem}

Theorem~\ref{th3}(a) establishes the first-order
approximation of Cook's distance for a large class of parametric
models for both dependent and independent data.
This leads to a substantial savings in computational time, since
it is computationally easier to calculate $ \mathbf{f}_{I}(\hat{\ve\theta
})$, $\mathbf{F}_{n}(\hat{\ve\theta})$ and $\mathbf{s}_{I}(\hat{\ve\theta})$
compared to $\operatorname{CD}(I)$.
Theorem~\ref{th3}(b) and (c) give an approximation of $\mathrm{E}[\operatorname
{CD}(I)|{\mathcal
M}]$ and $\mathrm{E}[\operatorname{CD}(I)|{\mathcal M}, \mathbf{Z}]$,
respectively. Generally, it is difficult to give a simple
approximation to
$\operatorname{Var}[\operatorname{CD}(I)|{\mathcal
M}]$ and $\operatorname{Var}[\operatorname{CD}(I)|{\mathcal M}, \mathbf{Z}]$, since it involves
the fourth moment of
$\mathbf{f}_{I}(\hat{\ve\theta})$, which does not have a simple form.

Based on Theorem~\ref{th3}, we can approximate the scaled
Cook's distance measures as follows.

Step (i). We generate a random sample $\mathbf{Y}^s$ from $p(\mathbf{Y}|\mathbf{
Z}, \hat{\ve\theta})$
and calculate $\widetilde{\operatorname{CD}}(I)$ based on the simulated sample~$\mathbf{Y}^s$ and fixed $\mathbf{Z}$,
denoted by~$\widetilde{\operatorname{CD}}(I)^{s}$. Explicitly, to calculate
$\widetilde{\operatorname{CD}}(I)^{s}$, we replace $\mathbf{Y}$ in
$ \mathbf{f}_{I}(\hat{\ve\theta})$, $\mathbf{F}_{n}(\hat{\ve\theta})$, and
$\mathbf{s}_{I}(\hat{\ve\theta})$ by~$\mathbf{Y}^s$. The computational burden
involved in computing
$\widetilde{\operatorname{CD}}(I)^{s}$
is very minor.

Compared to the exact computation of the scaled
Cook's distances,
we have avoided computing the maximum likelihood estimate of $\ve\theta
$ based on~$\mathbf{Y}^s$, which leads to great computational savings in computing
$\widetilde{\operatorname{CD}}(I)^{s}$
for large $S$, say \mbox{$S>100$}.
Theoretically, since $\hat{\ve\theta}$ is a consistent estimate of ${\ve
\theta}_* $,
$E[\widetilde{\operatorname{CD}}(I)|{\mathcal M}]$ is a consistent estimate of
$E[{\operatorname{CD}}(I)|{\mathcal M}]$.
Compared with reestimating $\hat{\ve\theta}^s$ for each~$\mathbf{Y}^s$,
a drawback of using $\hat
{\ve\theta}$ in calculating
$\widetilde{\operatorname{CD}}(I)^{s}$ is that $\widetilde{\operatorname{CD}}(I)^{s}$
does not account for the variability in $\hat{\ve\theta}$.
Similar arguments hold for the other
three quantities of ${\operatorname{CD}(I)}$.

Step (ii). By repeating Step (i) $S$ times, we can
use the empirical quantities of $\{\widetilde{\operatorname{CD}}(I)^{s}\dvtx  s=1,
\ldots, S\}$ to approximate
$E[\operatorname{CD}(I)|{\mathcal M}, \mathbf{Z}]$, $\operatorname{Std}[\operatorname
{CD}(I)|{\mathcal M}, \mathbf{Z}]$,
$Q_{\operatorname{CD}(I)}(0.5|{\mathcal M}, \mathbf{Z})$ and $\operatorname{Mstd}[\operatorname
{CD}(I)|{\mathcal M}, \mathbf{Z}]$. Subsequently, we can
approximate $\operatorname{CSCD}_1(I, \mathbf{Z})$ and $\operatorname{CSCD}_2(I, \mathbf{Z})$
and determine their magnitude based on $\widetilde{\operatorname{CD}}(I)^{s}$.

For instance,
let $\widehat{M}[\widetilde{\operatorname{CD}}(I)]$ and $\widehat{\operatorname
{Std}}[\widetilde{\operatorname{CD}}(I)]$ be, respectively, the sample mean and
standard deviation of $\{\widetilde{\operatorname{CD}}(I)^{s}\dvtx  s=1, \ldots, S\}
$. We calculate
\begin{eqnarray*}
\widetilde{\operatorname{CSCD}}_1(I, \mathbf{Z})&=&
\frac{\{\widetilde{\operatorname{CD}}(I)-\widehat{M}[\widetilde{\operatorname
{CD}}(I)]\}}{\widehat{\operatorname{Std}}[\widetilde{\operatorname{CD}}(I)]},\\
\widetilde{\operatorname{CSCD}}_1(I, \mathbf{Z})^s&=&
\frac{\{\widetilde{\operatorname{CD}}(I)^s-\widehat{M}[\widetilde{\operatorname
{CD}}(I)]\}}{\widehat{\operatorname{Std}}[\widetilde{\operatorname{CD}}(I)]}.
\end{eqnarray*}
We use $\widetilde{\operatorname{CSCD}}_1(I, \mathbf{Z})$ to approximate
$\operatorname{CSCD}_1(I,  \mathbf{Z})$ and then compare\break $\widetilde{\operatorname
{CSCD}}_1(I, \mathbf{Z})$ across different $I$ in order to determine
whether a specific subset $I$ is relatively influential or not.
Moreover, since $\widetilde{\operatorname{CSCD}}_1(\tilde I, \mathbf{Z})^s$ can be
regarded as
the ``true'' scaled Cook's distance when $p(\mathbf{Y}|\mathbf{Z}, \hat{\ve
\theta})$ is true,
we can either
compare $\widetilde{\operatorname{CSCD}}_1(I, \mathbf{Z})$
with $\widetilde{\operatorname{CSCD}}_1(\tilde I, \mathbf{Z})^s$ for all subsets
$\tilde I$ and $s$ or
compare $\widetilde{\operatorname{CSCD}}_1(I, \mathbf{Z})$
with $\widetilde{\operatorname{CSCD}}_1(I, \mathbf{Z})^s$ for all $s$.
Specifically, we calculate two probabilities as follows:
%
\begin{eqnarray}\label{ProbEq}
P_A(I, \mathbf{Z})&=& \sum_{s=1}^S \mathbf{1}\bigl( \widetilde{\operatorname{CSCD}}_1( I,
\mathbf{Z})^s \leq\widetilde{\operatorname{CSCD}}_1(I, \mathbf{Z})\bigr)/S, \\
P_B(I, \mathbf{Z})&=&\sum_{\tilde I}\sum_{s=1}^S \frac{ \mathbf{1}( \widetilde
{\operatorname{CSCD}}_1(\tilde I, \mathbf{Z})^s \leq\widetilde{\operatorname{CSCD}}_1(I,
\mathbf{Z}))}{S\times\#(\tilde I)},
\end{eqnarray}
where $\#(\tilde I)$ is the total number of all possible sets, and
$\mathbf{1}(\cdot)$ is an indicator function of a set.
We regard a subset $I$ as influential
if the value of $P_A(I, \mathbf{Z})$ [or $P_B(I, \mathbf{Z})$)] is relatively large.
Similarly, we can use the same strategy to
quantify the size of $\operatorname{CSCD}_2(I, \mathbf{Z})$, $\operatorname{SCD}_1(I)$ and
$\operatorname{SCD}_2(I)$.

Another issue is the accuracy of the first-order approximation
$\widetilde{\operatorname{CD}}(I)$ to the exact ${\operatorname{CD}}(I)$.
For relatively influential subsets, even though the accuracy of the first-order
approximation may be relatively low, $\widetilde{\operatorname{CD}}(I)$ can
easily pick out these influential points. Thus, for
diagnostic purposes, the first-order approximation may be more
effective at identifying influential subsets compared to
the true Cook's distance. We conduct simulation studies to investigate
the performance of the first-order approximation $\widetilde{\operatorname
{CD}}(I)$ relative to the
exact ${\operatorname{CD}}(I)$. Numerical comparisons are given in Section~\ref{sec3}.

We consider cluster deletion in generalized linear mixed models
(GLMM).

\begin{example}\label{ex2}
Consider a dataset, that is, composed of
a response $y_{ij}$, covariate vectors $\mathbf{x}_{ij} (p\times1) $
and $\mathbf{c}_{ij} (p_1\times1)$, for observations $j=1, \ldots,
m_i$ within clusters $i=1, \ldots, n$.
The GLMM assumes
that conditional on a $p_1\times1$ random variable $\mathbf{b}_i$,
$y_{ij}$ follows an exponential family distribution of the form~\cite{McCullagh-Nelder1989}
%
\begin{equation}\label{zl1}
p(y_{ij}|\mathbf{b}_i)=\exp\{a(\tau)^{-1}[y_{ij}\eta_{ij}-b(\eta_{ij})]+c(y_{ij},
\tau)\},
\end{equation}
where $\eta_{ij}=k(\mathbf{x}_{ij}^T{\ve\beta}+\mathbf{c}_{ij}^T\mathbf{b}_i) $ in
which ${\ve\beta}=({\ve\beta}_1, \ldots, {\ve\beta}_{p})^T$ and $k(\cdot
)$ is a
known continuously differentiable function. The distribution of $\mathbf{
b}_i$ is assumed to be $N(\mathbf{0}, \Sigma)$, where
$\Sigma=\Sigma({\ve\gamma})$ depends on a $p_2\times1$ vector
${\ve\gamma}$ of unknown variance components. In this case, we fix all
covariates $\mathbf{x}_{ij}$ and $\mathbf{c}_{ij}$ and all
$m_i$ and include them in $\mathbf{Z}$. For simplicity,
we fix $({\ve\gamma}, \tau)$ at an appropriate estimate $(\hat{\ve\gamma},
\hat\tau)$ throughout the example.

We focus here on cluster deletion in GLMMs. After some calculations,
the first-order approximation of $\operatorname{CD}(I_i)$ for deleting the
$i$th cluster is given by
%
\begin{equation}\label{GLMMeq1}
\widetilde{\operatorname{CD}}(I_i)= \partial_{\beta} \ell_i(\hat{\ve\beta})^T
[\mathbf{F}_{n}(\hat{\ve\beta})-\mathbf{f}_{i}(\hat{\ve\beta})]^{-1}
\mathbf{F}_{n}(\hat{\ve\beta}) [\mathbf{F}_{n}(\hat{\ve\beta})-\mathbf{
f}_{i}(\hat{\ve\beta})]^{-1}\partial_{\beta}
\ell_i(\hat{\ve\beta}),\hspace*{-40pt}
\end{equation}
where $I_i=\{(i, 1), \ldots, (i, m_i)\}$, $\ell_i({\ve\beta})$ is the
log-likelihood function for the $i$th cluster, $\mathbf{
f}_i({\ve\beta})=-\partial_{\beta}^2\ell_i({\ve\beta})$ and $\mathbf{
F}_n({\ve\beta})=\sum_{i=1}^n\mathbf{f}_i({\ve\beta})$. Note that
\[
\partial_{\beta} \ell_i(\hat{\ve\beta}) \approx\{\mathbf{I}_p-\mathbf{
f}_i(\hat{\ve\beta})[\mathbf{F}_n({\ve\beta}_*)]^{-1}\}
\partial_{\beta} \ell_i({\ve\beta}_*)+\mathbf{
f}_i(\hat{\ve\beta})[\mathbf{F}_n({\ve\beta}_*)]^{-1}\sum_{j\not=i}\partial
_{\beta}
\ell_j({\ve\beta}_*).
\]
Then, conditional on all the covariates and $\{m_1,
\ldots, m_n\}$ in $\mathbf{Z}$, we can show that $\mbox{E}[\widetilde{\operatorname
{CD}}(I_i)|{\mathcal M}, \mathbf{Z}]$ can be
approximated by $ \operatorname{tr}(\{\mbox{E}[\mathbf{F}_n (\hat{\ve\beta
})|{\mathcal M},
\mathbf{Z}]-\mbox{E}[\mathbf{f}_i (\hat{\ve\beta})| {\mathcal M}, \mathbf{Z}]\}
^{-1}\mbox{E}[\mathbf{
f}_i (\hat{\ve\beta})|{\mathcal M}, \mathbf{Z}])$ when ${\mathcal M}$
is true. Moreover, we may approximate $\operatorname{Var}[\widetilde{\operatorname
{CD}}(I_i)|{\mathcal M}, \mathbf{Z}]$ by using the fourth moment of
$\partial_{\beta} \ell_i({\ve\beta}_*)$.
It is not straightforward to approximate $Q_{\operatorname{CD}(I_i)}(0.5|{\mathcal
M}, \mathbf{Z})$ and
$ {\operatorname{Mstd}[\operatorname{CD}(I_i)|{\mathcal M}, \mathbf{Z}]}$.
Computationally, we employ the parametric bootstrap method described above
to approximate the conditionally
scaled Cook's distances $\operatorname{CSCD}_1(I_i, \mathbf{Z})$
and
$\operatorname{CSCD}_2(I_i, \mathbf{Z})$.
\end{example}

\section{Simulation studies and a real data example}\label{sec3}

In this section, we illustrate our methodology with simulated data
and a real data example. We also include some additional results in the
supplemental article~\cite{ZhuSizeSupp2012}.
The code along with its documentation for implementing our methodology
is available on the first author's website at
\texttt{\href{http://www.bios.unc.edu/research/bias/software.html}{http://www.bios.unc.edu/research/bias/}
\href{http://www.bios.unc.edu/research/bias/software.html}{software.html}}.

\subsection{Simulation studies}\label{sec3.1}

The goals of our simulations were to examine the finite sample
performance of Cook's distance
and the scaled Cook's distances and their first-order approximations
for detecting influential clusters in longitudinal data.
We generated 100 datasets from a linear mixed model.
Specifically, each dataset contains $n$ clusters.
For each cluster, the random effect $b_i$ was first
independently
generated from a $N(0, \sigma_b^2)$ distribution and then, given $b_i$,
the observations $y_{ij}$ $(j=1, \ldots, m_i; i=1,\ldots, n)$ were
independently generated as
$y_{ij}\sim N(\mathbf{x}_{ij}^T {\ve\beta} + b_i, \sigma_y^2)$ and the $m_{i}$
were randomly drawn from $\{1, \ldots, 5\}$. The covariates
$\mathbf{x}_{ij}$ were set as $(1,u_{i}, t_{ij})^T$, among which $t_{ij}$
represents time, and $u_i$ denotes a baseline covariate. Moreover,
$t_{ij}=\log(j)$ and the
$u_{i}$'s were independently generated from a $N(0,1)$ distribution.
For all 100 datasets, the responses were repeatedly simulated, whereas
we generated
the covariates and cluster sizes only once in order to fix the effect
of the covariates and cluster sizes on Cook's distance for each cluster.
The true value of $\ve\theta=({\ve\beta}^T,\sigma_b, \sigma_y)^T$ was
fixed at $(1,
1, 1, 1, 1)^T$. The sample size $n$ was set at $12$ to represent a
small number of clusters.

For each simulated dataset, we considered
the detection of influential clusters~\cite{Banerjee-Frees1997}.
We fit\vadjust{\goodbreak} the same linear mixed model and
used the expectation--maximization (EM) algorithm to calculate $\hat{\ve
\theta}$ and $\hat{\ve\theta}_{[I]}$ for each
cluster $I$.
We treated $(\sigma_b, \sigma_y)$ as nuisance parameters and ${\ve\beta
}$ as the parameter vector
of interest.
We calculated the degree of the perturbation $ {\mathcal P}(\{i\}
|{\mathcal M})$ for
deleting each subject $\{i\}$ while fixing the covariates, and then
we calculated the conditionally scaled Cook's distances and associated
quantities.
Let $\mathbf{x}_i$ be an $m_i \times 3$ matrix with the $j$th row being~$\mathbf{x}^T_{i,j}$.
It can be shown that for the case of fixed
covariates, we have
%
\begin{equation}
{\mathcal P}(\{i\}|{\mathcal M})= 0.5 \operatorname{tr}\{ \mathbf{x}_i^T R_i(\hat
{\ve\alpha})^{-1}\mathbf{x}_i E_\beta[({\ve\beta}-{\ve\beta}_*)({\ve\beta
}-{\ve\beta}_*)^T] \},
\end{equation}
where $E_\beta$ is taken with respect to $p({\ve\beta}| {\ve\beta}_* ,
G_{n\beta}^{-1})$ and $R_i(\ve\alpha)=\sigma_y^2\mathbf{I}_{m_i}+\sigma
_b^2 \mathbf{1}_{m_i}^{\otimes2}$, in which $\ve\alpha=(\sigma_b^2, \sigma
_y^2)^T$ and $\mathbf{1}_{m_i}$ is an
$m_i\times1$ vector with all elements equal to one. We set $G_{n\beta
}^{-1}=[\sum_{i=1}^n\mathbf{x}_i^TR_i(\hat{\ve\alpha})^{-1}\mathbf{x}_i]^{-1} $
and substituted $\ve\beta_*$ by $\hat{\ve\beta}$.\vspace*{1pt}


We carried out three experiments as follows. The first experiment was to
evaluate
the accuracy of the first-order approximation to $\operatorname{CD}(I)$.
The explicit expression of $\widetilde{\operatorname{CD}}(I)$ is given in
Example S2 of the supplementary document.
We considered two scenarios. In the first scenario, we directly
simulated 100 datasets from the above linear mixed model.
In the second scenario,
for each simulated dataset, we deleted
all the observations in clusters $n-1$ and $n$ and then reset $(m_{1},
b_{1})=(1, 4)$ and $(m_{n}, b_{n})=(5, 3)$ to generate
$y_{i, j}$ for $i=1, n$ and all $j$ according to the above
linear mixed model. Thus, the new first and $n$th
clusters can be regarded as influential clusters due to the extreme
values of $b_{1}$ and $b_n$.
Moreover,
the number of observations
in these two clusters is unbalanced.
We
calculated $\operatorname{CD}(I)$ and $\widetilde{\operatorname{CD}}(I)$, the average
$\operatorname{CD}(I)$, and the
biases
and standard errors of the differences $\operatorname{CD}(I)-\widetilde{\operatorname
{CD}}(I)$ for each cluster~$\{i\}$ (Table~\ref{tab1}).


Inspecting Table~\ref{tab1} reveals three findings as follows.
First, when no influential cluster is present in the first scenario,
the average
$\operatorname{CD}(I)$ is an increasing function of ${\mathcal P}(I|{\mathcal
M})$, whereas it is only positively proportional to
the cluster size $n(I)$ with a correlation coefficient of 0.83.
This result agrees with the results of Proposition~\ref{prop1}.
Second, in the second scenario,
the average
$\operatorname{CD}(I)$ for the true ``good'' clusters is positively
proportional to ${\mathcal P}(I|{\mathcal M})$ with a correlation
coefficient of 0.76,
while that for the influential clusters is associated with both
${\mathcal P}(I|{\mathcal M})$ and
the amount of influence that we introduced.
Third,
for the true ``good''
clusters, the first-order approximation is very accurate and leads to small
average biases and standard errors.
Even for the influential clusters, $\widetilde{\operatorname{CD}}(I)$ is
relatively close to $\operatorname{CD}(I)$.
For instance, for cluster $\{n\}$, the bias of 0.19 is relatively small
compared with 0.78, the mean of $\operatorname{CD}(\{n\})$.

In the second experiment, we considered the same two scenarios
as the first experiment. Specifically,
for each dataset, we approximated
$E[\operatorname{CD}(I)|{\mathcal M}, \mathbf{Z}]$ and $\operatorname{Std}[\operatorname
{CD}(I)|{\mathcal M}, \mathbf{Z}]$ by setting
$S=200$ and using their empirical values, and calculated
their first approximations $\widehat{M}[\widetilde{\operatorname{CD}}(I)]$ and
$\widehat{\operatorname{Std}}[\widetilde{\operatorname{CD}}(I)]$.
Across all 100 data sets, for each cluster $I$,
we computed the averages of
$E[\operatorname{CD}(I)|{\mathcal M}, \mathbf{Z}]$ and $\operatorname{Std}[\operatorname
{CD}(I)|{\mathcal M}, \mathbf{Z}]$, and the
biases
and standard errors of the differences $E[\operatorname{CD}(I)|{\mathcal M},
\mathbf{Z}]
- \widehat{M}[\widetilde{\operatorname{CD}}(I)]$
and $ \operatorname{Std}[\operatorname{CD}(I)|{\mathcal M}, \mathbf{Z}]-\widehat{\operatorname
{Std}}[\widetilde{\operatorname{CD}}(I)]$.

Table~\ref{tab1} shows the results for each scenario.
First, in both scenarios, the average
$E[\operatorname{CD}(I)|{\mathcal M}, \mathbf{Z}]$ is an increasing function of
${\mathcal P}(I|{\mathcal M})$, whereas it is only positively
proportional to
the cluster size $n(I)$ with a correlation coefficient (CC) of 0.80.
This is in agreement with the results of Proposition~\ref{prop1}.
The average of
$\operatorname{Std}[\operatorname{CD}(I)|{\mathcal M}, \mathbf{Z}]$ are positively
proportional to
$m_i$ (CC${}={}$0.76) and ${\mathcal P}(I|{\mathcal M})$ (CC${}={}$0.99).
Second, for all
clusters, the first-order approximations of $E[\operatorname{CD}(I)|{\mathcal
M}, \mathbf{Z}]$ and $\operatorname{Std}[\operatorname{CD}(I)|{\mathcal M}, \mathbf{Z}]$ are
very accurate and lead to small
average biases and standard errors.

\begin{table}
\tabcolsep=0pt
\caption{Selected results from simulation studies for $n=12$ and the
two scenarios:
$m_i$, ${\mathcal P}(\{i\}|{\mathcal M})$, M, SD, Mdif ($\times
10^{-2}$) and SDdif ($\times10^{-1}$) of the three quantities $\operatorname{CD}(I)$,
$E[\operatorname{CD}(I)|{\mathcal M}, \mathbf{Z}]$ and $\operatorname{Std}[\operatorname{CD}(I)|{\mathcal
M}, \mathbf{Z}]$.
$m_i$ denotes the cluster size of subset $\{i\}$;
${\mathcal P}(\{i\}|{\mathcal M})$ denotes the degree of perturbation;
M denotes the mean; SD denotes the standard deviation;
Mdif and SDdif, respectively, denote the mean and standard deviation
of the differences between each quantity and its first-order approximation.
In the first scenario, all observations were generated from the linear
mixed model, while in the second scenario, two clusters were
influential clusters and highlighted in bold.
For each case, 100 simulated
datasets were used. Results were sorted according to the degree of
perturbation for each cluster}\label{tab1}
{\fontsize{8}{10}\selectfont{
\begin{tabular*}{\textwidth}{@{\extracolsep{\fill}}lccd{2.2}cclccd{3.2}cc@{}}
\hline
& \multicolumn{5}{c}{\textbf{Scenario I}}&&\multicolumn{5}{c@{}}{\textbf{Scenario II}}
\\[-6pt]
& \multicolumn{5}{c}{\hrulefill}&&\multicolumn{5}{c@{}}{\hrulefill} \\
$\bolds{m_i}$& $\bolds{{\mathcal P}(\{i\}|{\mathcal M})}$ & \textbf{M} & \textbf{Mdif} & \textbf{SD} & \textbf{SDdif} &
$\bolds{m_i}$ & $\bolds{{\mathcal P}(\{i\}|{\mathcal M})}$ &\textbf{M} & \multicolumn{1}{c}{\textbf{Mdif}} &\textbf{SD} & \textbf{SDdif} \\
\hline
\multicolumn{12}{c}{$\operatorname{CD}(I)$} \\
1& 0.10& 0.11 & 0.01 & 0.09 & 0.03 & \textbf{1}& \textbf{0.08} &
\textbf{0.37} & \multicolumn{1}{c}{\phantom{00}\textbf{1.01}} & \textbf{0.18} & \textbf{0.18} \\
2& 0.11 & 0.12 & 0.32 & 0.12 & 0.15 & 2 & 0.11 & 0.10 & 0.08 & 0.09 &
0.12 \\
2 & 0.11 & 0.15 & 1.24 & 0.18 & 0.64 & 1 & 0.11 & 0.08 & 0.02 & 0.11 &
0.02 \\
2 & 0.13 & 0.18 & 0.87 & 0.19 & 0.36 & 2 & 0.13 & 0.13 & 0.08 & 0.12 &
0.12 \\
2 & 0.15 & 0.17 & 0.25 & 0.19 & 0.20 & 2 & 0.16 & 0.13 & -0.13 & 0.12
& 0.08 \\
3 & 0.16 & 0.23 & 0.55 & 0.19 & 0.50 & 2 & 0.20& 0.20 & 0.08 & 0.19 &
0.12 \\
2 & 0.19 & 0.26 & -0.02 & 0.32 & 0.25 & 3 & 0.23& 0.21 & -0.06 & 0.18
& 0.22 \\
3 & 0.22 & 0.34 & 2.97 & 0.35 & 0.99 & 4 & 0.25 & 0.23 & 0.37 & 0.23 &
0.26 \\
4 & 0.27 & 0.41 & 3.35 & 0.38 & 1.77& \textbf{5} & \textbf{0.28} & \textbf{
0.78} & \multicolumn{1}{c}{\phantom{0}\textbf{18.59}} & \textbf{0.61} & \textbf{4.71} \\
5 & 0.40 & 0.70 & 5.43 & 0.60 & 1.90 & 5 & 0.37 & 0.38 & 0.90 & 0.32 &
0.46 \\
4 & 0.57 & 1.15 & 1.57 & 1.29 & 1.73 & 5 & 0.54& 0.70 & 1.32 & 0.68 &
0.82 \\
5 & 0.60 & 1.21 & 3.62 & 1.49 & 1.62 & 4 & 0.56 & 0.65 & 1.06 & 0.69 &
0.54 \\[3pt]
\multicolumn{12}{c}{$E[\operatorname{CD}(I)|{\mathcal M}, \mathbf{Z}]$ } \\
1 & 0.10 & 0.12 & 0.22 & 0.02 & 0.05 & \textbf{1} & \textbf{0.08} &
\textbf{0.09} & \multicolumn{1}{c}{\phantom{00}\textbf{0.43}} & \textbf{0.01} & \textbf{0.04} \\
2 & 0.11 & 0.12 & 0.41 & 0.01 & 0.03 & 2 & 0.11 & 0.12 & 0.45 & 0.02 &
0.04 \\
2 & 0.11 & 0.13 & 0.46 & 0.02 & 0.04 & 1 & 0.11 & 0.13 & 0.09 & 0.02 &
0.03 \\
2 & 0.12& 0.15 & 0.40 & 0.02 & 0.07 & 2 & 0.13 & 0.15 & 0.38 & 0.02 &
0.04 \\
2 & 0.15 & 0.17 & 0.34 & 0.03 & 0.08 & 2 & 0.16 & 0.18 & 0.26 & 0.02 &
0.04 \\
3 & 0.16 & 0.18 & 0.77 & 0.02 & 0.08 & 2 & 0.20 & 0.23 & 0.12 & 0.03 &
0.05 \\
2 & 0.19 & 0.22 & 0.21 & 0.04 & 0.09 & 3 & 0.23 & 0.27 & 0.46 & 0.03 &
0.07 \\
3 & 0.22 & 0.26 & 0.62 & 0.04 & 0.09 & 4 & 0.25 & 0.29 & 1.13 & 0.03&
0.13 \\
4 & 0.26 & 0.32 & 1.63 & 0.03 & 0.15 & \textbf{5} & \textbf{0.28} &
\textbf{0.36} & \multicolumn{1}{c}{\phantom{00}\textbf{1.94}} & \textbf{0.04}& \textbf{0.18} \\
5 & 0.40& 0.55 & 2.58 & 0.07 & 0.29 & 5 & 0.37 & 0.48 & 1.86 & 0.05 &
0.18 \\
4 & 0.57 & 0.97 & 2.21 & 0.12 & 0.21 &5 & {0.53} & {0.82} & 4.26 &
{0.10} & {0.34} \\
5 & 0.60 & 1.03 & 5.87 & 0.16 & 0.99 & 4 & 0.56 & 0.93 & 1.64 & 0.11 &
0.17 \\[3pt]
\multicolumn{12}{c}{$\operatorname{Std}[\operatorname{CD}(I)|{\mathcal M}, \mathbf{Z}]$ } \\
1 & 0.10 & 0.18 & 1.48 & 0.04 & 0.20 & \textbf{1} & \textbf{0.08} &
\textbf{0.13} & \multicolumn{1}{c}{\phantom{00}\textbf{1.05}} & \textbf{0.04} & \textbf{0.22} \\
2 & 0.11 & 0.14 & 1.16 & 0.03 & 0.10 & 2 & 0.11 & 0.14 & 1.18 & 0.03 &
0.12 \\
2 & 0.11 & 0.15 & 1.37 & 0.03 & 0.16 & 1 & 0.11 & 0.18 & 0.78 & 0.04 &
0.10 \\
2 & 0.13 & 0.18 & 1.72 & 0.05 & 0.35 & 2 & 0.13 & 0.18 & 1.15 & 0.03 &
0.13 \\
2 & 0.15 & 0.21 & 2.02 & 0.05 & 0.25 & 2 & 0.16 & 0.23 & 1.28 & 0.04 &
0.14 \\
3 & 0.16 & 0.19 & 2.05 & 0.03 & 0.25 & 2 & 0.20 & 0.30 & 1.07 & 0.06 &
0.16 \\
2 & 0.19 & 0.29 & 2.36 & 0.07 & 0.24 & 3 & 0.23 & 0.31 & 1.72 & 0.06 &
0.22 \\
3 & 0.22 & 0.30 & 2.55 & 0.07 & 0.32 & 4 & 0.25 & 0.30 & 1.96 & 0.05 &
0.42 \\
4 & 0.26 & 0.35 & 2.84 & 0.06 & 0.39 & \textbf{5} & \textbf{0.28} &
\textbf{0.39} & \multicolumn{1}{c}{\phantom{00}\textbf{4.06}} & \textbf{0.09} & \textbf{0.66} \\
5 & 0.40 & 0.58 & 2.13 & 0.11 & 0.71 & 5 & 0.37 & 0.50 & 2.67 & 0.09 &
0.52 \\
4 & 0.57 & 1.16 & 1.17 & 0.18 & 0.55 & {5} & {0.53} & {0.89} & 0.60 &
{0.14} & {0.68} \\
5 & 0.60 & 1.14 & -4.18 & 0.25 & 2.29 & 4 & 0.56 & 1.13 & 0.94 & 0.21 &
0.41 \\
\hline
\end{tabular*}}}
\end{table}

The third experiment was to
examine the finite sample performance of Cook's distance
and the scaled Cook's distances for detecting influential clusters in
longitudinal data.
We considered two scenarios. In the first scenario,
for each of the 100 simulated datasets, we deleted
all the observations in cluster $n$ and then reset $m_{n}=1$ and varied
$b_n$ from 0.6 to 6.0 to generate
$y_{n, j}$ according to the above
linear mixed model. The second scenario is almost the same as the first
scenario, except that we reset $m_n=10$.
Note that when the value of $b_n$ is relatively large, for example,
$b_n=2.5$, the $n$th cluster is an influential cluster, whereas the
$n$th cluster is not influential
for small $b_n$.
A good case-deletion measure should detect the $n$th cluster as truly
influential for large $b_n$, whereas it does not for small $b_n$.
For each data set, we approximated
$\operatorname{CSCD}_1(I, \mathbf{Z})$, $\operatorname{CSCD}_2(I, \mathbf{Z})$,
$\widetilde{\operatorname{CSCD}}_1(I, \mathbf{Z})$ and $\widetilde{\operatorname
{CSCD}}_2(I, \mathbf{Z})$ by setting $S=100$.
Subsequently, we calculated $P_A(I, \mathbf{Z})$ and $P_B(I, \mathbf{Z})$
in (\ref{ProbEq}) and
$P_C(I, \mathbf{Z})=\sum_{I\not=\{n\}} \mathbf{1}(\operatorname{CD}(I)\leq\operatorname
{CD}(\{n\}))/(n-1)$.
Finally, across all 100 datasets, we calculated the averages and
standard errors of
all diagnostic measures for the $n$th cluster for each scenario.

\begin{figure}

\includegraphics{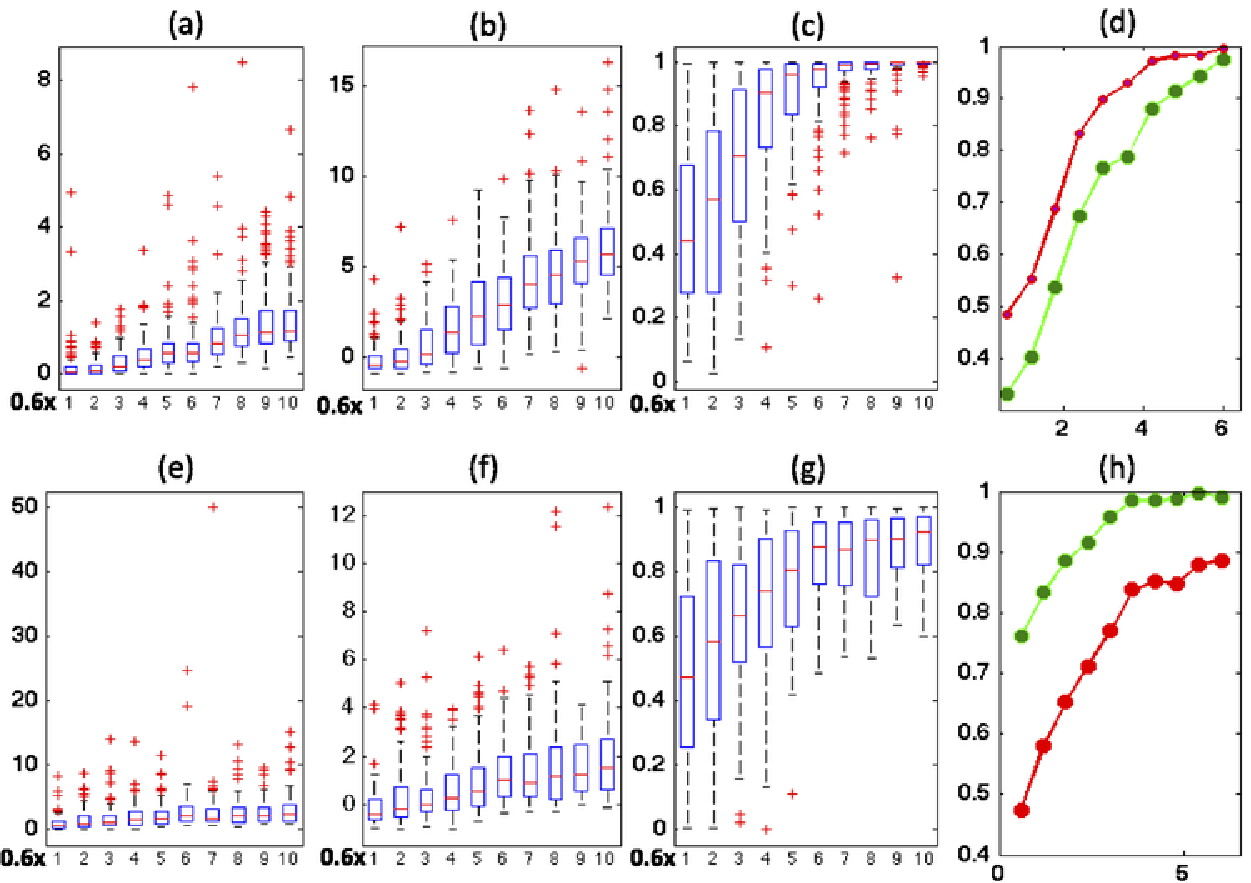}

\caption{Simulation results from 100 datasets simulated from a linear
mixed model in the two scenarios.
The first row corresponds to the first scenario, in which $m_{12}=1$
and~$b_{12}$ varies from 0.6 to~6.0. The second row corresponds to the
second scenario, in which $m_{12}=10$ and $b_{12}$ varies from~0.6 to~6.0.
Panels \textup{(a)} and \textup{(e)} show the box plots of Cook's distances as a function
of~$b_{12}$;
panels \textup{(b)} and \textup{(f)} show the box plots of $\operatorname{CSCD}_1(I, \mathbf{Z})$ as
a function of~$b_{12}$;
panels \textup{(c)} and \textup{(g)} show the box plots of $P_B(I, \mathbf{Z})$ as a
function of~$b_{12}$;
panels \textup{(d)} and \textup{(h)} show the mean curve of $P_B(I, \mathbf{Z})$ based on
$\operatorname{CSCD}_1(I, \mathbf{Z})$ (red line) and the mean curve of
$P_C(I, \mathbf{Z})$ based on
$\operatorname{CD}(I)$ (green line) as functions of~$b_{12}$.}\label{fig1}
\end{figure}

Inspecting Figure~\ref{fig1} reveals some findings as follows. First,
deleting the $n$th cluster with 10 observations causes a larger effect
than that with 1 observation [Figure~\ref{fig1}(a) and (e), (d) and (h)].
As expected, the distributions of $\operatorname{CD}(\{n\})$ and $\widetilde
{\operatorname{CSCD}}_1(I, \mathbf{Z})$ shift up as $b_n$ increases [Figure~\ref{fig1}(a),
(b), (e) and (f)].
Second, in the first scenario, $\operatorname{CD}(\{n\})$ is stochastically
smaller than most other $\operatorname{CD}(I)$s, when the value of $b_n$ is
relatively small [Figure~\ref{fig1}(d)]. However,
in the second scenario,
$\operatorname{CD}(\{n\})$ is stochastically larger than most other $\operatorname
{CD}(I)$s [Figure~\ref{fig1}(h)] even for small values of $b_n$.
Specifically, when $m_n=1$, the average $P_C(\{n\}, \mathbf{Z})$ is
smaller than 0.4 as $b_n=0.6$ and $b_n=1.2$, whereas when $m_n=10$, the average
$P_C(\{n\}, \mathbf{Z})$ is higher than 0.75 even as $b_n=0.6$.
In contrast, in the two scenarios,
the value of $P_B(\{n\}, \mathbf{Z})$ is close to 0.5 as $b_n=0.6$ [Figure~\ref{fig1}(d) and (h)].
It indicates that the cluster size does not have a big effect on the
distribution of $\widetilde{\operatorname{CSCD}}_1(I, \mathbf{Z})$ [Figure~\ref{fig1}(c) and~(g)].

\subsection{Yale infant growth data}\label{sec3.2}

The Yale infant growth data were collected to study whether cocaine
exposure during pregnancy may lead to the maltreatment of infants after
birth, such as physical and sexual abuse. A total of $298$ children
were recruited
from two subject groups (cocaine exposed group and unexposed group).
One feature of this dataset is that the number of observations per children
$m_i$ varies significantly from $2$ to $30$
\cite{Wasserman-Leventhal1993,Stier-etal1993}. The total number of
data points is $\sum_{i=1}^nm_i=3176$.
Following Zhang~\cite{Zhang1999}, we considered two
linear mixed models given by $ y_{i, j}=\mathbf{x}_{i, j}^T{\ve\beta
}+\varepsilon_{i, j},
$ where $y_{i, j}$ is the weight (in kilograms) of the $j$th visit from
the $i$th subject,
$\mathbf{x}_{i, j}=(1, d_{i, j}, (d_{i, j}-120)^+, (d_{i, j}-200)^+,
(g_{i}-28)^+,d_{i, j}(g_{i}-28)^+, (d_{i, j}-60)^+(g_{i}-28)^+, (d_{i,
j}-490)^+(g_i-28)^+, s_id_{i, j},
s_i(d_{i, j}-120)^+)^T$, in which $d_{i, j}$ and $g_i$ (days) are the
age of visit and
gestational age, respectively, and $s_i$ is the indicator for gender.
In addition, we assumed
$\ve\varepsilon_i=(\varepsilon_{i,1}, \ldots, \varepsilon_{i, m_i})^T\sim
N_{m_{i}}(\mathbf{0}, R_i(\ve\alpha))$, where $\ve\alpha$ is a vector of
unknown parameters in $R_i(\ve\alpha)$.
We first considered $R_i(\ve\alpha)=\alpha_0\mathbf{I}_{m_i}+\alpha_1
\mathbf{1}_{m_i}^{\otimes2}$. We refer to this model as model $M_1$.
Then, it is assumed that variance and autocorrelation parameters are,
respectively, given by
$V(d)=\exp(\alpha_0+\alpha_1 d+\alpha_2 d^2+\alpha_3 d^3)$ and $\rho
(l)=\alpha_4+\alpha_5 l$, where $l$ is the lag between two visits.
We refer to this model as model $M_2$.

\begin{figure}

\includegraphics{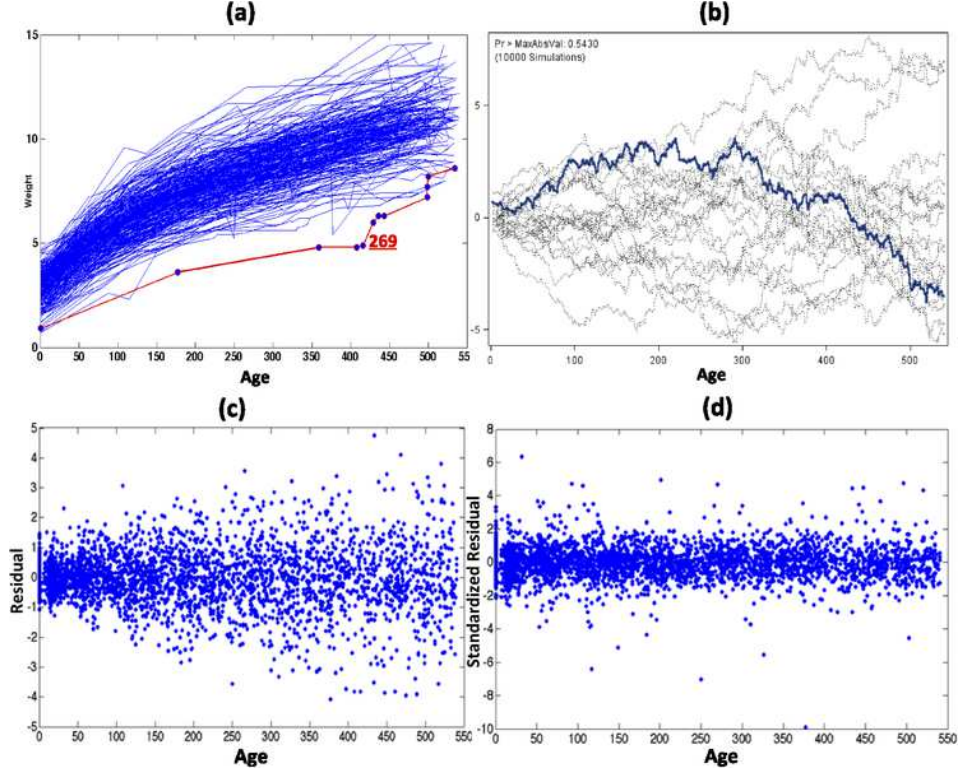}

\caption{Yale infant growth data. Panel \textup{(a)} presents the
line plot of infant weight against age, in which the observations of
subject 269 are highlighted; panel \textup{(b)} shows the cumulative residual
curve versus
age, in which the observed cumulative residual curve is highlighted in
blue; and
panels \textup{(c)} and \textup{(d)}, respectively, present age versus raw residual and
age versus standardized residual for cluster deletion.}\label{fig2}
\end{figure}

We systematically examined the key assumptions of models $M_1$ and
$M_2$ as follows.

(i) We presented a cumulative residual plot and calculated the cumulative
sums of residuals over the age of the visit to test $E[y_{i, j}|\mathbf{
x}_{i, j}]=\mathbf{x}_{i, j}^T{\ve\beta}$~\cite{Lin2002}, whose $p$-value
is greater than 0.543. It may suggest that the mean structure is
reasonable. The cumulative residual plot is given in Figure~\ref{fig2}(b).

(ii) For model $M_1$, inspecting the plot of raw residuals $r_{i,
j}=y_{i,j}-\mathbf{x}_{i, j}^T\hat{\ve\beta}$ against age in Figure~\ref{fig2}(c)
reveals that
the variance of the raw residuals
appears to increase with the age of visit. As pointed by Zhang~\cite{Zhang1999}, it may be more sensible to use model $M_2$.
Let $\tilde\mathbf{r}_i=(\tilde r_{i, 1}, \ldots, \tilde r_{i,
m_i})^T=R_i(\hat{\ve\alpha})^{-1/2}\mathbf{r}_i$ be the vector of
standardized residuals of $M_2$, where
$\mathbf{r}_i=(r_{i, 1}, \ldots, r_{i, m_i})^T$. The standardized
residuals under $M_2$ do not have any
apparent structure as age increases [Figure~\ref{fig2}(d)].

(iii) Under each model,
we calculated $\operatorname{CD}(I)$
for each child~\cite{Banerjee-Frees1997}.
We treated~${\ve\beta}$ as parameters of interest and all elements of
$\ve\alpha$ as nuisance parameters. For model~$M_1$,
we
obtained a strong Pearson correlation of 0.363 between Cook's distance
and the
cluster size.
This indicates that the bigger the cluster size, the larger the Cook's
distance measure. Figure 4(b) highlights the top ten influential
subjects.
Compared with model~$M_1$, we observed similar findings by using $\operatorname
{CD}(I)$ under model $M_2$, which were omitted for space limitations.

There are several difficulties in using
Cook's distance under both models $M_1$ and $M_2$
\cite{Preisser-Qaqish1996,Christensen-etal92,Banerjee-Frees1997,Banerjee1998}.
First, cluster size varies
significantly across children, and deleting a
larger cluster may have a higher probability of having a larger
influence as discussed in Section~\ref{sec2.3}.
For instance, we observe $(m_{285},
\operatorname{CD}(\{285\}))=(8, 0.738)$ and $(m_{274}, \operatorname{CD}(\{274\}))=(22,
1.163)$. A~larger $\operatorname{CD}(\{274\})$ can be
caused by a larger $m_{274}=22$ and/or
influential subject 274, among others.
Since $m_{274}$ is much larger than $m_{285}$, it is
difficult to claim that subject $274$ is more influential than
subject ${285}$.
Second, there is no rule for determining whether a specific subject is
influential relative to the fitted model. Specifically,
it is unclear whether the subjects with larger $\operatorname{CD}(\{i\})$ are
truly influential or not.
Third, inspecting Cook's distance solely does not seem to
delineate the potential misspecification of the covariance structure
under model $M_1$.
We will address these three difficulties by using the new case-deletion
measures.

\begin{figure}

\includegraphics{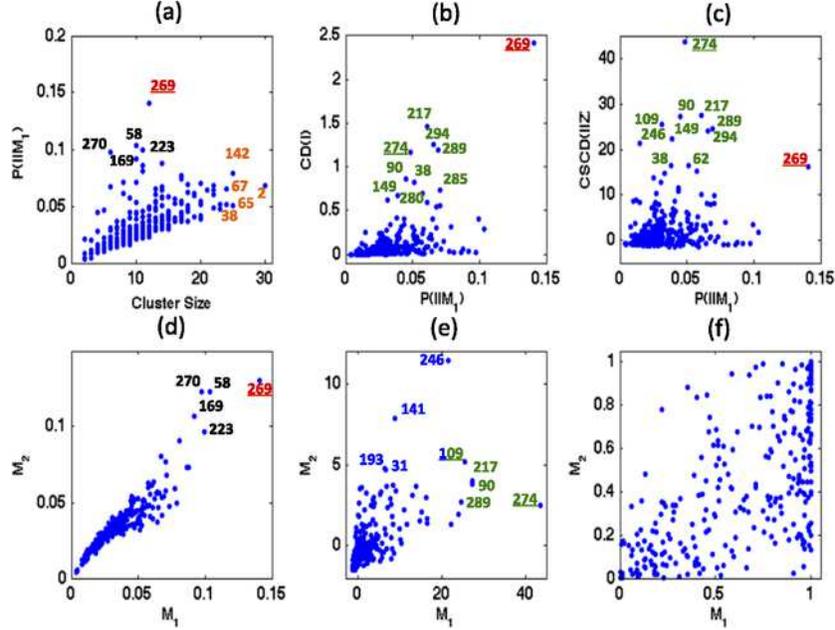}

\caption{Yale infant growth data. Panel \textup{(a)} shows $m_i$
versus ${\mathcal P}(I|{\mathcal M}_1)$, in which the ten subjects with
the largest values of degree of perturbation or cluster size are
highlighted; panel~\textup{(b)} shows ${\mathcal P}(I|{\mathcal M}_1)$
versus $\operatorname{CD}(I)$, in which the top ten influential subjects are
highlighted; panel \textup{(c)} shows ${\mathcal P}(I|{\mathcal M}_1)$
versus $\operatorname{CSCD}_1(I, \mathbf{Z})$, in which the top eleven
influential subjects are highlighted; and panels \textup{(d)}, \textup{(e)} and \textup{(f)},
respectively, show
${\mathcal P}(I|{\mathcal M})$,
$\operatorname{CSCD}_1(I, \mathbf{Z})$ and $P_B(I, \mathbf{Z})$
for models ${\mathcal M}_1$ and ${\mathcal M}_2$.}\label{fig3}
\end{figure}

(iv) Under each model,
we calculated $ {\mathcal P}(\{i\}|{\mathcal M})$ for
deleting each subject~$\{i\}$ for fixed covariates, and then
we calculated the conditionally scaled Cook's distances and associated
quantities.
We then used 1000 bootstrap samples to approximate
${\operatorname{CSCD}}_1(I, \mathbf{Z})$ and ${\operatorname{CSCD}}_2(I, \mathbf{Z})$.
Subsequently, we calculated $P_A(I, \mathbf{Z})$ and $P_B(I, \mathbf{Z})$
in (\ref{ProbEq}).

We observed several findings. First, under model $M_1$, we
observed a strong positive correlation between $ {\mathcal P}(\{i\}
|{\mathcal M})$ and $m_i$ [Figure~\ref{fig3}(a)].
Second, even though $m_{269}=12$ is moderate, subject
$269$ has the largest degree of perturbation. Inspecting the raw data
in Figure~\ref{fig2}(a) reveals that subject 269 is of older age during visits
compared with other subjects.
Third, we also
observed a strong positive correlation between $ {\mathcal P}(\{i\}
|{\mathcal M})$ and Cook's distance [Figure~\ref{fig3}(b)], which may indicate their stochastic
relationship as discussed in Section~\ref{sec2.3}.
Fourth, we observed a positive correlation between Cook's
distance and the conditionally scaled Cook's distance
[Figure~\ref{fig3}(b) and (c)], but their levels of influence for the same subject
are quite different. For instance, the magnitude of
$\operatorname{CSCD}_1(\{269\}, \mathbf{Z})$ is only moderate, whereas $\operatorname
{CD}_1(\{269\}, \mathbf{Z})$ is the highest one.
We observed similar findings under model~${\mathcal M}_2$
and presented some findings in Figure~\ref{fig3}(d) and (e).

We used
$P_B(I, \mathbf{Z})$ to quantify whether a specific subject is influential
relative to the fitted model ${\mathcal M}_1$ [Figure~\ref{fig3}(f)].
For instance, since $\operatorname{CD}(\{246\})=0.253$, it is unclear whether
subject 246 is
influential or not according to $\operatorname{CD}$,
whereas we have $\operatorname{CSCD}_1(\{ 246\}, \mathbf{Z})=21.443$ and $P_B(\{
246\}, \mathbf{Z})=1.0 $. Thus, subject 246 is really influential after
eliminating the effect of the cluster size.
Moreover, it is difficult to compare the influential levels of
subjects $274$ and $285$ using $\operatorname{CD}$. All of the conditionally
scaled Cook's distances and associated quantities suggest that
subject $274$ is more influential than subject $285$ after eliminating
the degree of perturbation difference.
We observed similar findings under model ${\mathcal M}_2$
and omitted them due to space limitations. See Figure~\ref{fig3}(d) and (e) for details.

We compared the goodness of fit of models ${\mathcal M}_1$ and
${\mathcal M}_2$ to the data by using the proposed case-deletion measures.
First, inspecting Figure~\ref{fig3}(d) reveals a strong similarity between the
degrees of perturbation under models ${\mathcal M}_1$ and ${\mathcal
M}_2$ for all subjects.
Second, by using the conditionally scaled Cook's distance,
we observed different levels of influence for the same subject
under~${\mathcal M}_1$ and ${\mathcal M}_2$.
For instance, $\operatorname{CSCD}_1(I, \mathbf{Z})$ identifies
subjects $246, 141, 109, 193$ and $31$ as the top five influential
subjects under ${\mathcal M}_1$, whereas it identifies
subjects $274, 217, 90, 109$ and $289$ as the top ones under
${\mathcal M}_2$.
Finally, examining $P_B(I, \mathbf{Z})$ reveals a large percentage of
influential points for model ${\mathcal M}_1$, but a small percentage
of influential points for model ${\mathcal M}_2$; see Figure~\ref{fig3}(f) for
details. This may indicate that model ${\mathcal M}_2$ outperforms
model ${\mathcal M}_1$. Furthermore, although
we may
develop goodness-of-fit statistics based on the scaled Cook's
distances and show that model ${\mathcal M}_2$ outperforms model~${\mathcal M}_1$,
this will be a topic of our future research.

In summary, the use of the new case-deletion measures provides new
insights in real data analysis.
First, ${\mathcal P}(I|{\mathcal M})$ explicitly quantifies the degree
of perturbation introduced by deleting each subject.
Second, $\operatorname{CSCD}_k(I, \mathbf{Z})$ for $k=1, 2$ explicitly account
for the degree of perturbation for each subject.
Third, $P_B(I, \mathbf{Z})$ allows us to quantify whether a specific
subject is influential relative to the fitted model.
Fourth, inspecting $P_B(I, \mathbf{Z})$ and $\operatorname{CSCD}_k(I, \mathbf{Z})$ may
delineate the potential misspecification of the covariance structure
under model $M_1$.

\section{Discussion}\label{sec4}
We have introduced a new quantity to quantify the degree of
perturbation and examined its properties.
We have
used stochastic ordering to quantify the
relationship between the degree of the perturbation and the magnitude of
Cook's distance.
We have developed several scaled Cook's distances to address the
fundamental issue of deletion diagnostics in general parametric
models. We have shown that the scaled
Cook's distances provide important information about the relative
influential level of each subset.
Future work includes developing goodness-of-fit statistics based on the
scaled Cook's distances,
developing Bayesian analogs
to the scaled Cook's distances, and developing user-friendly \texttt{R}
code for implementing our proposed measures in various models, such as
survival models and
models with missing covariate data.

\begin{appendix}
\section*{Appendix}\label{app}

The following assumptions are needed to facilitate the technical
details, although they are not the weakest possible conditions.
Because we develop all results for general parametric models, we
only assume several high-level assumptions as follows.

\begin{assumption}\label{assa2}
$\hat{{\ve\theta}}_{[I]}$ for any $I$
is a consistent estimate
of ${\ve\theta}_*\in\Theta$.
\end{assumption}

\begin{assumption}\label{assa3}All $p(\mathbf{Y}_{[I]}| {\ve\theta})$ are
three times continuously differentiable on $\Theta$ and satisfy
\begin{eqnarray*}
\log p\bigl( \mathbf{Y}_{[I]}| {\ve\theta}\bigr)&=&\log p\bigl(\mathbf{Y}_{[I]}| {\ve\theta}_*
\bigr)+\Delta({\ve\theta})^T
J_{n, {[I]}}({\ve\theta}_* )\\
&&{}-
0.5\Delta({\ve\theta})^T\mathbf{F}_{n, {[I]}}({\ve\theta}_* )\Delta({\ve
\theta})+R_{[I]}({\ve\theta}),
\end{eqnarray*}
in which $|R_{[I]}({\ve\theta})|=o_p(1)$ uniformly for all ${\ve\theta}
\in
B({\ve\theta}_*, \delta_0n^{-1/2})=\{{\ve\theta}\dvtx
\sqrt{n}\Vert {\ve\theta}-{\ve\theta}_* \Vert \leq\delta_0\}$, where
$\Delta({\ve\theta})={\ve\theta}-{\ve\theta}_* $, $J_{n,
{[I]}}({\ve\theta})=\partial_\theta\log p(\mathbf{Y}_{[I]}| {\ve\theta})$ and
$\mathbf{F}_{n, {[I]}}({\ve\theta}_* )=\partial_\theta^2 \log p(\mathbf{
Y}_{[I]}| {\ve\theta})$.
\end{assumption}

\begin{assumption}\label{assa4}
For any $I$ and $\mathbf{Z}$,
$\sup_{ \theta\in B({\theta}_* , n^{-1/2}\delta_0)}
n^{-1/2}J_{n, {[I]}}({\ve\theta})=O_p(1),$
\begin{eqnarray*}
\sup_{\theta\in B({\theta}_* ,
n^{-1/2}\delta_0)} \bigl\Vert \mathbf{F}_{n, [I]}({\ve\theta})-E[\mathbf{
F}_{I}({\ve\theta})|{\mathcal M}, \mathbf{Z}]\bigr\Vert &=&O_p\bigl(\sqrt{n}\bigr), \\
\sup_{ \theta, \theta'\in B({\theta}_* ,
n^{-1/2}\delta_0)} n^{-1}\bigl\Vert \mathbf{F}_{n, [I]}({\ve\theta})-\mathbf{F}_{n,
[I]}(\ve\theta')\bigr\Vert &=&o_p(1),
\end{eqnarray*}
and
$ 0< \inf_{\theta\in B({ \theta}_*,
\delta_0n^{-1/2})}\lambda_{\min}(n^{-1} \mathbf{F}_{n,
[I]}({\ve\theta}))\leq\sup_{\theta\in B({ \theta}_*,
\delta_0n^{-1/2})}\lambda_{\max}(n^{-1} \times \mathbf{F}_{n,
[I]}({\ve\theta}))<\infty. $
\end{assumption}

\begin{assumption}\label{assa5}
For any set $I$ and $\mathbf{Z}$,
\begin{eqnarray*}
\sup_{\theta\in B({ \theta}_* , n^{-1/2}\delta_0)}
J_{ {I}}({\ve\theta})&=&O_p\bigl(\sqrt{n(I)}\bigr), \\
 \sup_{\theta\in B({ \theta
}_* ,
n^{-1/2}\delta_0)} \Vert \mathbf{f}_{ I}({\ve\theta})\Vert &=&O_p(n(I)),
\\
 \sup_{\theta\in B({ \theta}_* , n^{-1/2}\delta_0)} \Vert \mathbf{f}_{
I}({\ve\theta})-E[\mathbf{f}_{I}({\ve\theta})|{\mathcal M}, \mathbf{
Z}]\Vert &=&O_p\bigl(\sqrt{n(I)}\bigr).
\end{eqnarray*}
\end{assumption}

\begin{remarks*}
Assumptions~\ref{assa2}--\ref{assa5} are very general
conditions and are generalizations of some higher level conditions
for the extremum estimator, such as the maximum likelihood estimate,
given in Andrews~\cite{Andrews1999}. Assumption~\ref{assa2} assumes that the parameter
estimates with and without deleting the observations in the subset $I$
are consistent. Assumption~\ref{assa3} assumes that the log-likelihood functions
for any~$I$ and $\mathbf{ Y}_{[I]}$ admit a second-order Taylor's series expansion
in a small neighborhood of~${\ve\theta}_*$. Assumptions~\ref{assa4} and~\ref{assa5} are
standard assumptions to ensure that the first- and second-order
derivatives of $p(\mathbf{Y}_{[I]}|{\ve\theta})$ and $p(\mathbf{Y}_{I}|\mathbf{
Y}_{[I]}, {\ve\theta})$ have appropriate rates of $n$ and $n_I$~\cite{Andrews1999,Zhu-Zhang2006}.
Sufficient conditions of Assumptions~\ref{assa2}--\ref{assa5} have
been extensively discussed in the literature~\cite{Andrews1999,Zhu-Zhang2006}.
\end{remarks*}

\begin{pf*}{Proof of Theorem \protect\ref{th1}}
(P.a) directly follows from the
Jensen inequality, (\ref{DPeq1}) and (\ref{DPeq2}).
For (P.b), if $I$ is an empty set, then $\operatorname{KL}(\mathbf{Y}, {\ve\theta
}| I)\equiv0$ and thus ${\mathcal P}(I|{\mathcal M})=0$.
On the other hand, if ${\mathcal P}(I|{\mathcal M})=0$, then $\operatorname
{KL}(\mathbf{Y}, {\ve\theta}| I)\equiv0$ for almost every $\ve\theta$.
Thus, by using
the Jensen inequality, we have $p(\mathbf{Y}_{I}|\mathbf{Y}_{[I]},\allowbreak {\ve\theta
})\equiv p(\mathbf{Y}_{I}|\mathbf{Y}_{[I]}, {\ve\theta}_* )$ for all $\ve
\theta\in\Theta$.
Based on the identifiability condition, we know that~$I$ must be an
empty set.
Let $I_{1\cdot2}=I_1-I_2$. It is easy to show that
\[
p\bigl(\mathbf{Y}_{I_1}|\mathbf{Y}_{[I_1]}, {\ve\theta}\bigr)= p\bigl(\mathbf{Y}_{I_2}, \mathbf{
Y}_{I_{1\cdot2}}|\mathbf{Y}_{[I_1]}, {\ve\theta}\bigr)
= p\bigl(\mathbf{Y}_{I_2}|\mathbf{Y}_{[I_2]}, {\ve\theta}\bigr) p\bigl(\mathbf{Y}_{[I_2]}|
\mathbf{Y}_{[I_1]}, {\ve\theta}\bigr).
\]
Thus, by substituting the above equation into (\ref{DPeq1}), we have
%
\begin{eqnarray}\label{Th1Eq1}
{\mathcal P}(I_1|{\mathcal M})&=& {\mathcal P}(I_2|{\mathcal M})
\nonumber\hspace*{-35pt}
\\[-8pt]
\\[-8pt]
\nonumber
&&{}+\int
p({\ve\theta}| {\ve\theta}_* , \Sigma_{n*}) p(\mathbf{Y}|{\ve\theta}) \log
\biggl(
\frac{p(\mathbf{Y}_{[I_2]}| \mathbf{Y}_{[I_1]}, {\ve\theta}) }{p(\mathbf{
Y}_{[I_2]}| \mathbf{Y}_{[I_1]}, {\ve\theta}_* ) }
\biggr) \,d{\ve\theta} \,d\mathbf{Y},\hspace*{-35pt}
\end{eqnarray}
in which the second term on the right-hand side can be written as
\begin{eqnarray*}
&&\int p({\ve\theta}| {\ve\theta}_* , \Sigma_{n*}) p\bigl(\mathbf{Y}_{I_2}| \mathbf{
Y}_{[I_2]}, {\ve\theta}\bigr)\\
&&\qquad{}\times\biggl\{ \int p\bigl(\mathbf{Y}_{[I_2]}|{\ve\theta}\bigr) \log
\biggl(
\frac{p(\mathbf{Y}_{[I_2]}| \mathbf{Y}_{[I_1]}, {\ve\theta}) }{p(\mathbf{
Y}_{[I_2]}| \mathbf{Y}_{[I_1]}, {\ve\theta}_* ) }
\biggr) \,d\mathbf{Y}_{[I_2]}\biggr\} \,d{\ve\theta} \,d\mathbf{Y}_{I_2}\geq0,
\end{eqnarray*}
which yields (P.c).
Based on the assumption of (P.d), we know that
\[
p\bigl(\mathbf{Y}_{[I_2]}| \mathbf{Y}_{[I_1]}, {\ve\theta}\bigr)=p\bigl(\mathbf{Y}_{I_{1\cdot
2}}| \mathbf{Y}_{[I_1]}, {\ve\theta}\bigr)=p\bigl(\mathbf{Y}_{I_{1\cdot2}}| \mathbf{
Y}_{[I_{1\cdot2}]}, {\ve\theta}\bigr)
\]
for all $\ve\theta$.
Thus,
the second term on the right-hand side of (\ref{Th1Eq1}) reduces to
${\mathcal P}(I_{1\cdot2}|{\mathcal M})$, which finishes the proof of (P.d).
\end{pf*}

\begin{pf*}{Proof of Theorem \protect\ref{th2}} (a) Let $I_3=I_1\setminus I_2$,
$I_1$ is a union of two disjoint sets~$I_3$ and $I_2$. Without loss
of generality, $H_{I_1}$ can be decomposed as
\[
H_{I_1}=\mathbf{X}_{I_1}(\mathbf{X}^T\mathbf{X})^{-1}\mathbf{X}_{I_1}^T=\pmatrix{
\mathbf{X}_{I_2}(\mathbf{X}^T\mathbf{X})^{-1}\mathbf{X}_{I_2}^T & \mathbf{
X}_{I_2}(\mathbf{X}^T\mathbf{X})^{-1}\mathbf{X}_{I_3}^T
\vspace*{2pt}\cr
\mathbf{X}_{I_3}(\mathbf{X}^T\mathbf{X})^{-1}\mathbf{X}_{I_2}^T & \mathbf{
X}_{I_3}(\mathbf{X}^T\mathbf{X})^{-1}\mathbf{X}_{I_3}^T}.
\]
Let $\lambda_{1,1}\geq\cdots\geq\lambda_{1, n(I_1)}\geq0$ and
$\lambda_{2, 1}\geq\cdots\geq\lambda_{2, n(I_2)}\geq0$ be the ordered
eigenvalues of $H_{I_1}$ and $H_{I_2}$, respectively, where
$n(I_k)$ denotes the number of observations in $I_k$ for $k=1, 2$. It
follows from Wielandt's
eigenvalue inequality~\cite{Eaton-Tyler1991} that $\lambda_{1,
l}\geq\lambda_{2, l}$ for all $l=1, \ldots, n(I_2)$. For $k=1, 2$, we
define $\Gamma_{k}\Lambda_{k}\Gamma_{k}^T$ as the spectral
decomposition of $H_{I_k}$ and $\mathbf{h}_k=(\mathbf{
I}_{n(I_k)}-\Lambda_k)^{-1/2}\Gamma_k^T\hat\mathbf{e}_{I_k}=(h_{k,1},
\ldots, h_{k, n(I_k)})^T$, where $\Gamma_k$ is an orthnormal matrix
and $\Lambda_k=\operatorname{diag}(\lambda_{k, 1}, \ldots, \lambda_{k,
n(I_k)})$. It can be shown that for $k=1, 2$,
\[
\mathbf{h}_k\sim N\bigl(\mathbf{0},
\sigma^2 \mathbf{I}_{n(I_k)}\bigr)
\quad\mbox{and} \quad  \operatorname{CD}(I_k)=\frac{1}{\hat\sigma^2} \sum_{j=1}^{n(I_k)}
\frac{\lambda_{k, j}}{1-\lambda_{k, j}}h_{k, j}^2.
\]
Since $f(x)=x/(1-x)$ is an increasing function of $x\in(0, 1)$, this
completes the proof of Theorem~\ref{th2}(a).

Note that $ \operatorname{CD}(I)=({\hat\sigma^2})^{-1} \sum_{j=1}^{n(I)}
{\lambda_{j}}{(1-\lambda_{j})}^{-1}h_{j}^2,$ where
the $\lambda_j$ are the eigenvalues of $H_I$ and $\mathbf{h}=(h_1, \ldots,
h_{n(I)})^T\sim N(\mathbf{0}, \sigma^2 \mathbf{I}_{n(I)})$. Moreover, the
distribution of $\lambda$
is uniquely determined by $H_{I}$. Combining $\mathbf{h}\sim N(\mathbf{
0}, \sigma^2 \mathbf{I}_{n(I)})$ with the assumptions of Theorem~\ref{th2}(b)
yields that $\operatorname{CD}(I)$ and $\operatorname{CD}(I')$ follow the same
distribution when $n(I)=n(I')$.
Furthermore,
we can always choose an $I_2'$ such that $n(I_2')=n(I_2)$ and
$I_1\subset I_2'$. Following arguments in Theorem~\ref{th2}(a), we can then
complete the
proof of Theorem~\ref{th2}(b).\vadjust{\goodbreak}
\end{pf*}


\begin{pf*}{Proof of Theorem \protect\ref{th3}}
(a) It follows from a Taylor series expansion and Assumption~\ref{assa3} that
\[
\partial_\theta\log p\bigl(\mathbf{Y}_{[I]}|
\hat{\ve\theta}_{[I]}\bigr)=\mathbf{0}=\partial_\theta\log p\bigl(\mathbf{Y}_{[I]}|
\hat{\ve\theta}\bigr)+\partial_\theta^2 \log p\bigl(\mathbf{Y}_{[I]}|
\tilde{\ve\theta}\bigr)\bigl(\hat{\ve\theta}_{[I]}-\hat{\ve\theta}\bigr),
\]
where $\tilde{\ve\theta}=t\hat{\ve\theta}_{[I]}+(1-t)\hat{\ve\theta}$
for $t\in[0,
1]$. Combining this with Assumption~\ref{assa4} and the fact that $\partial
_\theta\log
p(\mathbf{Y}| \hat{\ve\theta})=\partial_\theta\log p(\mathbf{Y}_{[I]}|
\hat{\ve\theta})+\partial_\theta\log p(\mathbf{Y}_{I}|\mathbf{Y}_{[I]},
\hat{\ve\theta})=\mathbf{0}$, we get
%
\begin{eqnarray}
\hat{\ve\theta}_{[I]}-\hat{\ve\theta}&=&\bigl[-\partial_\theta^2 \log
p\bigl(\mathbf{
Y}_{[I]}| \hat{\ve\theta}\bigr)\bigr]^{-1}\partial_\theta\log p\bigl(\mathbf{Y}_{[I]}|
\hat{\ve\theta}\bigr)[1+o_p(1)]\nonumber\\
&=&-\bigl[-\partial_\theta^2 \log p\bigl(\mathbf{
Y}_{[I]}| \hat{\ve\theta}\bigr)\bigr]^{-1}\partial_\theta\log p\bigl(\mathbf{Y}_{I}|\mathbf{
Y}_{[I]},
\hat{\ve\theta}\bigr)[1+o_p(1)]. \label{Th4Eq2}
\end{eqnarray}
Substituting (\ref{Th4Eq2}) into
$\operatorname{CD}(I)=(\hat{\ve\theta}_{[I]}-\hat{\ve\theta})^T\mathbf{
F}_n(\hat{\ve\theta})(\hat{\ve\theta}_{[I]}-\hat{\ve\theta})$ completes
the proof
of Theorem~\ref{th3}(a).

(b) It follows from Assumptions~\ref{assa2}--\ref{assa4} that
\begin{eqnarray*}
\hat{\ve\theta}-{\ve\theta}_* &=&\mathbf{F}_n({\ve\theta}_* )^{-1}\partial
_\theta\log p(\mathbf{
Y}| {\ve\theta}_* )[1+o_p(1)]
\\
&=&\mathbf{F}_n({\ve\theta}_* )^{-1}\bigl[\partial_\theta\log
p\bigl(\mathbf{Y}_{[I]}| {\ve\theta}_* \bigr)+\partial_\theta\log p\bigl(\mathbf{Y}_I|\mathbf{
Y}_{[I]}, {\ve\theta}_* \bigr)\bigr][1+o_p(1)].
\end{eqnarray*}
Let $J_I({\ve\theta})=\partial_\theta\log p(\mathbf{Y}_{I}|\mathbf{Y}_{[I]},
{\ve\theta})$. Using a Taylor series expansion along with Assumptions
\ref{assa4} and~\ref{assa5},
we get
%
\begin{eqnarray}
J_I(\hat{\ve\theta})
&=& J_I({\ve\theta}_* )- \mathbf{s}_I({\ve\theta}_* ) (\hat{\ve\theta}-{\ve
\theta}_* )[1+o_p(1)]\nonumber\\
&=& J_I({\ve\theta}_* )-E[\mathbf{s}_I({\ve\theta}_* )|{\mathcal M}](\hat
{\ve\theta}-{\ve\theta}_* )[1+o_p(1)]
\nonumber
\\[-8pt]
\\[-8pt]
\nonumber
&=&
\bigl(\{\mathbf{I}_p-
E[\mathbf{s}_I({\ve\theta})|{\mathcal M}]\mathbf{F}_n({\ve\theta}_* )^{-1}\}
J_I({\ve\theta}_* )\\
&&{}-  E[\mathbf{s}_I({\ve\theta})|{\mathcal M}]\mathbf{F}_n({\ve
\theta}_* )^{-1}\partial_\theta\log p\bigl(\mathbf{Y}_{[I]}|
{\ve\theta}_* \bigr)\bigr)[1+o_p(1)].\nonumber
\end{eqnarray}
Since $E[J_I({\ve\theta}_* )\partial_\theta\log p(\mathbf{Y}_{[I]}|
{\ve\theta}_* )|{\mathcal M}]=\mathbf{0}$,
\begin{eqnarray*}
&&E[J_I(\hat{\ve\theta})J_I(\hat{\ve\theta})^T|{\mathcal M}]\\
&&\qquad=E[\mathbf{
s}_I({\ve\theta}_* )|{\mathcal M}]\mathbf{F}_n({\ve\theta}_* )^{-1}\{\mathbf{
F}_n({\ve\theta}_* )-E[\mathbf{s}_I({\ve\theta}_* )|{\mathcal M}]\}[1+o_p(1)].
\end{eqnarray*}
It follows
from Assumption~\ref{assa4} that for $\ve\theta$ in a neighborhood of
${\ve\theta}_* $, $\mathbf{F}_n({\ve\theta})$ and $\mathbf{F}_n({\ve\theta}_*
)-\mathbf{f}_I({\ve\theta})$ can be replaced by $E[\mathbf{F}_n({\ve\theta
})|{\mathcal M}]$
and $E[\mathbf{F}_n({\ve\theta}_* )-\mathbf{f}_I({\ve\theta})|{\mathcal M}]$,
respectively, which completes the proof of Theorem~\ref{th3}(b).

(c) Similarly to Theorem~\ref{th3}(b), we can prove Theorem~\ref{th3}(c).
\end{pf*}
\end{appendix}


\begin{supplement}[id=suppA]
\sname{Supplement to ``Perturbation and scaled Cook's distance''}
\slink[doi,text={10.1214/12-AOS978SUPP}]{10.1214/12-AOS978SUPP} 
\sdatatype{.pdf}
\sfilename{aos978\_supp.pdf}
\sdescription{We include two theoretical examples and
additional results obtained from the Monte Carlo simulation studies and
real data analysis.}
\end{supplement}

\section*{Acknowledgments}
We thank the Editor Peter B\"{u}hlmann, the Associate Editor and two anonymous referees for
valuable suggestions, which have greatly helped to improve our presentation.

%

%

\printaddresses

\end{document}